\documentclass{new_tlp}
\usepackage{amssymb}
\usepackage{amstext}
\usepackage{amsmath}
\usepackage{verbatim}
\usepackage{mathptmx}
\usepackage{graphicx}
\usepackage{times}
\usepackage{helvet}   
\usepackage{courier}
\usepackage{epsfig}
\usepackage{color}
\usepackage{url}
\usepackage{multicol}
\usepackage[stable]{footmisc}
  
\newtheorem{example}{Example}
\newtheorem{definition}{Definition}
\newtheorem{proposition}{Proposition}
\newtheorem{theorem}{Theorem}

\newcommand{\nop}[1]{}

\newcommand{\B}{\mbox{$\cal B$}}
\newcommand{\T}{\mbox{$T$}}
\newcommand{\TSM}{\mbox{$TSM$}}

\newcommand{\M}{\mbox{$\cal M$}}
\newcommand{\D}{\mbox{$D$}}

\newcommand{\LP}{\mbox{$LP$}}
\newcommand{\MP}{\mbox{$MP$}}

\newcommand{\IC}{\mbox{$IC$}}
\newcommand{\PS}{\mbox{$PS$}}
\newcommand{\MM}{\mbox{$MM$}}

\newcommand{\SM}{\mbox{$SM$}}

\newcommand{\PP}{\mbox{$P$}}
\newcommand{\WM}{\mbox{$WM$}}
\newcommand{\PWM}{\mbox{$MaxMinWM$}}
\newcommand{\PSM}{\mbox{$PSM$}}

\newcommand{\NP}{\mbox{$NP$}}

\newcommand{\MINWM}{\mbox{$MinWM$}}
\newcommand{\MAXWM}{\mbox{$MaxWM$}}

\def\-{\mbox{-}}
\def\+{\mbox{+}}
\def\<{\mbox{$\langle$}}
\def\>{\mbox{$\rangle$}}
\def\:{\mbox{\hspace*{-0.04mm}$:$}\hspace*{-0.04mm}}

\def\cldl{\vspace{-0.4cm}\[\bf\begin{array}{lcl}}
\def\eldl{\end{array}\]\rm}

\long\def\comment#1{}

\def\chn#1{\textcolor{black}{#1}}
\def\ch#1{\textcolor{black}{#1}}
\def\chg#1{\textcolor{black}{#1}}
\def\chr#1{\textcolor{black}{#1}}
\def\chrr#1{\textcolor{black}{#1}}
\begin{document}

  \title[A Logic Framework for P2P Deductive Databases]
        {A Logic Framework for P2P Deductive Databases}

  \author[Caroprese, Zumpano]
         {Luciano Caroprese\\
         \email{l.caroprese@dimes.unical.it}
\and 
Ester Zumpano\\
 \email{e.zumpano@dimes.unical.it}\\
 \\
  University of Calabria - 
  Department of Computer, Modelling, Electronics and Systems Engineering \\
   Via P. Bucci, Cubo 42C, 5th floor, 87036 Rende (CS), Italy
}

\maketitle

\vspace*{2mm}
\begin{abstract}
	\vspace*{2mm}
\ch{
This paper presents a logic framework for modeling the interaction
among deductive databases in a P2P (Peer to Peer) environment. \\
Each peer joining a P2P system \emph{provides or imports
	data} from its neighbors by using a set of \emph{mapping rules},
i.e. a set of semantic correspondences to a set of peers belonging
to the same environment. By using mapping rules, as soon as it
enters the system, a peer can participate and access all data
available in its neighborhood, and through its neighborhood it
becomes accessible to all the other peers in the system. 
A query can be posed to any peer in the system and the
answer is computed by using locally stored data and all the
information that can be consistently imported from the neighborhood. \\
Two different types of mapping rules are defined:
mapping rules allowing to import a maximal set of atoms not leading to inconsistency (called
\emph{maximal mapping rules}) and mapping rules allowing to import a 
minimal set of atoms needed to restore consistency (called \emph{minimal mapping rules}).
Implicitly, the use of 
maximal mapping rules states \emph{it is preferable to import as long as no inconsistencies arise}; whereas the use of 
minimal mapping rules states that \emph{it is preferable not to import unless a inconsistency exists}. \\
The paper presents three different declarative semantics of a P2P system: \\
(i) the \emph{Max Weak Model Semantics}, in which mapping rules are used to import \emph{as much knowledge as possible} from a peer's neighborhood without violating local integrity constraints; \\
(ii) the \emph{Min Weak Model  Semantics}, in which the  P2P system can be locally inconsistent and the information provided by the neighbors is  used to restore consistency, that is to only integrate the missing portion of a correct,
but incomplete database; \\
(iii) the  \emph{Max-Min Weak Model Semantics} that unifies the previous two different perspectives captured by the Max Weak Model Semantics and Min Weak Model Semantics. 
This last semantics allows to characterize each peer in the neighborhood as a resource used either to enrich (integrate) or to fix (repair) the knowledge,
so as to define a kind of \emph{integrate\-repair} strategy for
each peer.
For each semantics, the paper also introduces an equivalent and alternative characterization, obtained by rewriting each mapping rule into prioritized rules so as to model a P2P system as a prioritized logic program. \\
Finally, results about the computational complexity of P2P logic queries, are investigated by considering \emph{brave} and
\emph{cautious} reasoning. \\
\emph{Under consideration in Theory and Practice of Logic Programming (TPLP).}
\nop{----------------------------
More specifically, by considering analogous results on stable model
semantics for prioritized logic programs, the paper proves that for disjunction-free
($\vee-free$) prioritized programs
deciding whether an interpretation 
$M$ is a preferred weak model of $\PS$ is 
$co\NP$ complete; 
deciding whether
an atom is \emph{true} in some preferred model is
$\Sigma_2^p$-complete, whereas deciding whether an atom is
\emph{true} in every preferred model is $\Pi_2^p$-complete. Moreover, the paper also provides results on the existence of a preferred wek model showing that the problem 
is in $\Sigma_2^p$.
---------------------------------}
}
	
\end{abstract}


\section{Introduction}
Data exchange consists in sharing data from a source schema to a target schema according to specifications fixed by  source-to-target constraints \cite{DBLP:journals/tcs/FaginKMP05,DBLP:journals/tods/FaginKP05,DBLP:journals/tods/FuxmanKMT06}.
This challenging topic is closely  related to  data integration and consistent query answering  \cite{DBLP:conf/pods/Lenzerini02,DBLP:journals/tkde/GrecoGZ03,DBLP:conf/pods/ArenasBC99,DBLP:conf/sigmod/LeoneGILTEFFGRLLRKNS05,AreBer*99,DBLP:conf/pods/CaliLR03,DBLP:journals/is/CaliCGL04}. 
Data integration is one of the most fundamental processes in
intelligent systems, from individuals to societies. At the present,
the most important application of data integration is  any
form of P2P interaction and cooperation. 
Ideally, in P2P systems there is no selection, but integration of the
valuable contributions of every participant.

In a \emph{Peer Data Managment System}, PDMS, a number of peers interact and exchange data. 
More specifically, each peer joining a P2P system uses a set of
\emph{mapping rules}, i.e. a set of semantic correspondences to a
set of peers belonging to the same environment, to both
\emph{provide or import data} from its neighbors. Therefore, in a
P2P system the entry of a new source, \emph{peer}, is extremely
simple as it just requires the definition of the mapping rules. By
using mapping rules, as soon as it enters the system, a peer can
participate and access all data available in its neighborhood, and
through its neighborhood it becomes accessible to all other
peers in the system.

\noindent The possibility for the users of sharing knowledge from a
large number of informative sources, has enabled the development of
new methods for data integration easily usable for processing
distributed and autonomous data.

\noindent Due to this, there have been several proposals
which consider the integration of information and the computation of
queries in an open ended network of distributed peers
\cite{Bern-etal02,BertossiB04,Calv*04,DBLP:conf/dbisp2p/CalvaneseDGLR03,Franconi*04a} as well as the
problem of schema mediation ~\cite{Halevy*03,MadAhl*03,DBLP:journals/vldb/HalevyIST05}, query answering  and query optimization in P2P
environments~\cite{DBLP:conf/pods/AbiteboulD98,TatHal04,GribbleHIRS01,DBLP:journals/tcs/FaginKMP05}.

Previously proposed approaches investigate the data
integration problem in a P2P system by considering each peer as
locally consistent. Therefore, the introduction of inconsistency is only caused by the operation of importing data from other peers.
These approaches assume that for each peer it is preferable to
import as much knowledge as possible.

Our previous works, in the context of P2P data integration, follow
this direction.

\ch{In \cite{CarGre*06a,Car*06b,SUM2007,ISMIS2008,FOIKS2012}
it is adopted the classical idea that a peer imports maximal sets of atoms.
More specifically, the interaction among deductive databases in a P2P system has been
modeled by importing maximal sets of
atoms not violating integrity constraints, that is \emph{maximal
sets of atoms that allow the peer to enrich its knowledge while
preventing inconsistency anomalies}.
The following examples will clarify the perspective used by 
\emph{maximal mapping rules} 
to import in each peer maximal sets of atoms not violating integrity constraints.
}
\ch{
\begin{example}\label{Motivating-example-Max}
\begin{figure}[h]
	\centering
	\includegraphics[width=0.9\textwidth]{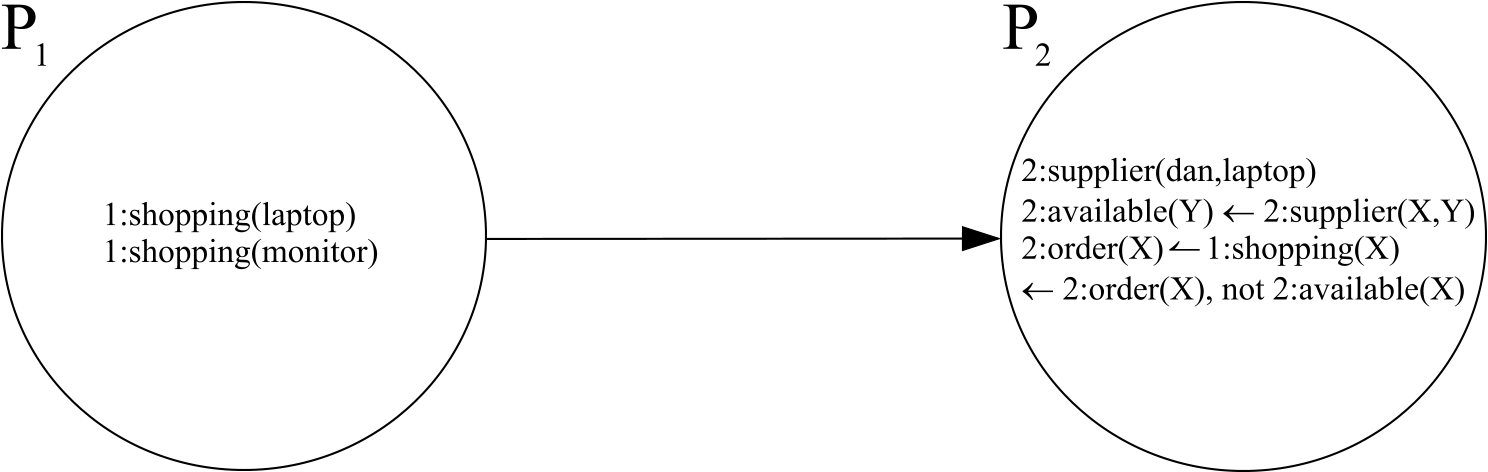}
	\caption{A P2P System with maximal mapping rules}
	\label{fig-Motivating-example-Max}
\end{figure}
Consider the P2P system depicted in Figure \ref{fig-Motivating-example-Max}.
\begin{itemize}
	\item
	Peer $\PP_1$ stores information about products that should be
	ordered. It contains the facts:  \\ $1\:shopping(laptop)$ and $1\:shopping(monitor)$. 
	The special syntax used for a fact -- its first part is the \textit{peer identifier} -- will be formally presented in Section \ref{P2PSystem}.\\
	\item
	Peer $\PP_2$ contains:
	\begin{itemize} 
		\item the fact $2\:supplier(dan,$ $laptop)$,
		whose meaning is \emph{`Dan is a supplier of laptops'};
		\item
		the maximal mapping rule,
		$2\:order(X) \leftharpoonup 1\:shopping(X)$, whose  precise syntax and
		semantics will be formally defined in Section \ref{P2PSystem}.
		Intuitively,  this rule allows to import as many orders as possible
		from the relation shopping of $\PP_1$ into the relation order of $\PP_2$. In fact,  it states that if $1\:shopping(X)$ is \emph{true} in the source peer $\PP_1$,  the atom $2\:order(X)$ can
		be imported in the target peer $\PP_2$ (that is $2\:order(X)$ is \emph{true} in the
		target peer) only if it does not imply the violation of some integrity constraints;
		\item
		the rule  $2\:available(Y) \leftarrow 2\:supplier(X,Y)$
		stating that a product $Y$ is available if there is a supplier $X$ of $Y$
		\item
		the integrity constraint $\leftarrow 2\:order(X), \  not \ 2\:available(X)$,  
		stating that the order of a device cannot exist if it is not available.
	\end{itemize} 
\end{itemize}
Intuitively, peer $\PP_1$ provides two facts, but  the maximal set of them that $\PP_2$ can import, using the mapping rule, is $\{2\:order(laptop)\}$. The fact $2\:order(monitor)$ cannot be imported as it would violate the integrity constraint; in fact, no supplier of the device \textit{monitor} exists.
\end{example}
}

Besides the basic classical idea followed in the previous example, a different 
perspective could be argued. Often, in real world P2P systems, peers use the available
import mechanisms to extract knowledge from the rest of the system
only if this knowledge is strictly needed to repair an inconsistent
local database. The work in \cite{CarZum12} stems from this
different perspective. A peer can be locally inconsistent and it can use the
information provided by its neighbors in order to
restore consistency, that is to only integrate the missing portion
of a correct, but incomplete database. Then, an inconsistent peer,
in the interaction with different peers, just imports the
information allowing to restore consistency,  that is \emph{minimal
sets of atoms allowing the peer to enrich its knowledge so as to restore  inconsistency anomalies}.

\noindent The following example will intuitively clarify this perspective.

\ch{
\begin{example}\label{Motivating-Example-minimal}
\begin{figure}[h]
	\centering
	\includegraphics[width=0.9\textwidth]{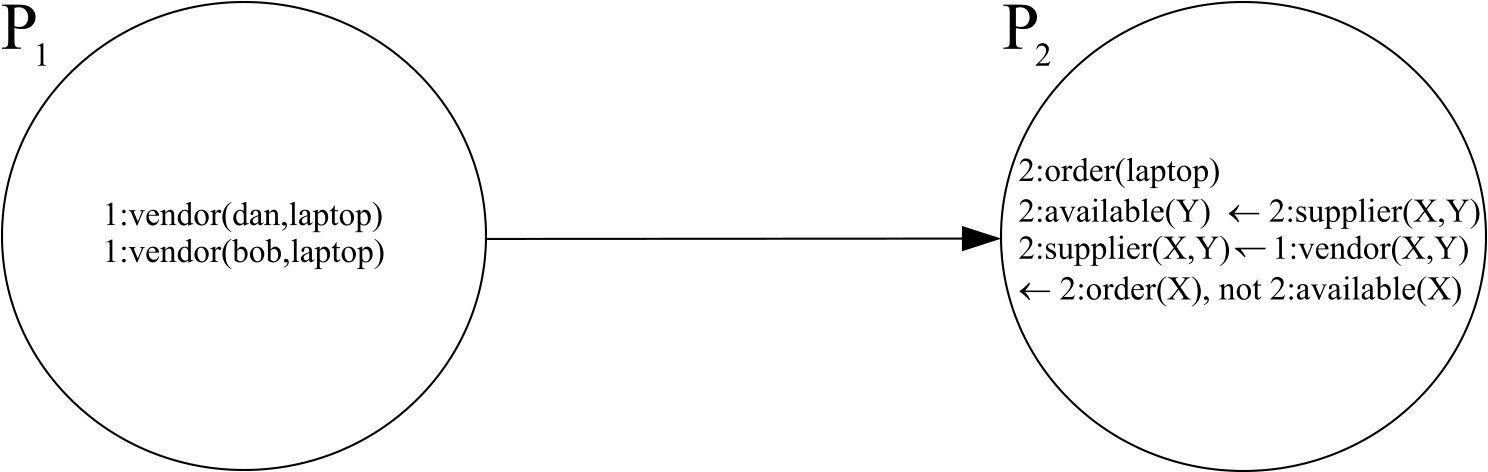}
	\caption{A P2P System with minimal mapping rules}
	\label{fig-Motivating-example-Min}
\end{figure}
Consider the P2P system depicted in Figure \ref{fig-Motivating-example-Min}. 
It consists of the following two peers:
\begin{itemize}
	\item
	Peer $\PP_1$ stores information about vendors of devices and contains the following  facts:   $1\:vendor$ $(dan, laptop)$,
	whose meaning is \emph{`Dan is a vendor of laptops'},  and $1\:vendor(bob,$ $laptop)$, whose meaning is
	\emph{`Bob is a vendor of laptops'}. \\
	\item
	Peer $\PP_2$ contains:
	\begin{itemize} 
		\item
		the fact $2\:order(laptop)$, stating that the order of a laptop exists;
		\item
		the minimal mapping rule
		$2\:supplier(X,Y) \leftharpoondown 1\:vendor(X,Y)$,  whose  precise syntax and
		semantics will be formally defined in Section \ref{P2PSystem}. 
		The rule is used to import tuples from the relation \emph{vendor} of $\PP_1$ into the relation \emph{supplier} of $\PP_2$. Intuitively,   the rule states that if 
		$1\:vendor(X,Y)$ is \emph{true} in the source peer the atom $2\:supplier(X,Y)$ can
		be imported in the target peer (that is $2\:supplier(X,Y)$ is \emph{true} in the
		target peer) only if it implies the
		satisfaction of some constraints that otherwise would be violated;
		\item
		the standard rule $2\:available(Y)\leftarrow 2\:supplier(X,Y)$, stating that a device $Y$ is
		available if there is a supplier $X$ of $Y$, 
		\item
		the integrity constraint $\leftarrow 2\:order(X), \ not \ 2\:available(X)$,
		stating that the order of a device cannot exist if it is not available.
	\end{itemize}
\end{itemize}
\noindent
Peer $\PP_2$ is inconsistent. The integrity constraint is violated as  the ordered device \emph{laptop}
is not available (there is no supplier of laptops).  The device $laptop$  needs to be provided by a supplier. Therefore, 
$\PP_2$ `needs' to import from its neighbors minimal sets of atoms in order to restore consistency. The intuition is that either  $1\:vendor(dan,$ $ laptop)$
or $1\:vendor(bob, laptop)$ can be imported into $\PP_2$ to satisfy the constraint (but not both).
~\hfill~$\Box$
\end{example}
}

The two concepts proposed in \cite{CarGre*06a,Car*06b,SUM2007,ISMIS2008,FOIKS2012} and in \cite{CarZum12} can be merged. 
The basic idea is that a peer of a P2P system can use each neighbor
to extract either as much knowledge as possible (i.e. to integrate its knowledge) or
just the portion that is strictly needed (i.e. to repair the knowledge of the system). This unified framework defines a sort of \emph{integrate\-repair} strategy.

The following example  will intuitively clarify our
perspective and will be used as a running example in the rest of the
paper.

\ch{
\begin{example}
\label{Motivating-example-Max-Min}
\begin{figure}[h]
	\centering
	\includegraphics[width=0.9\textwidth]{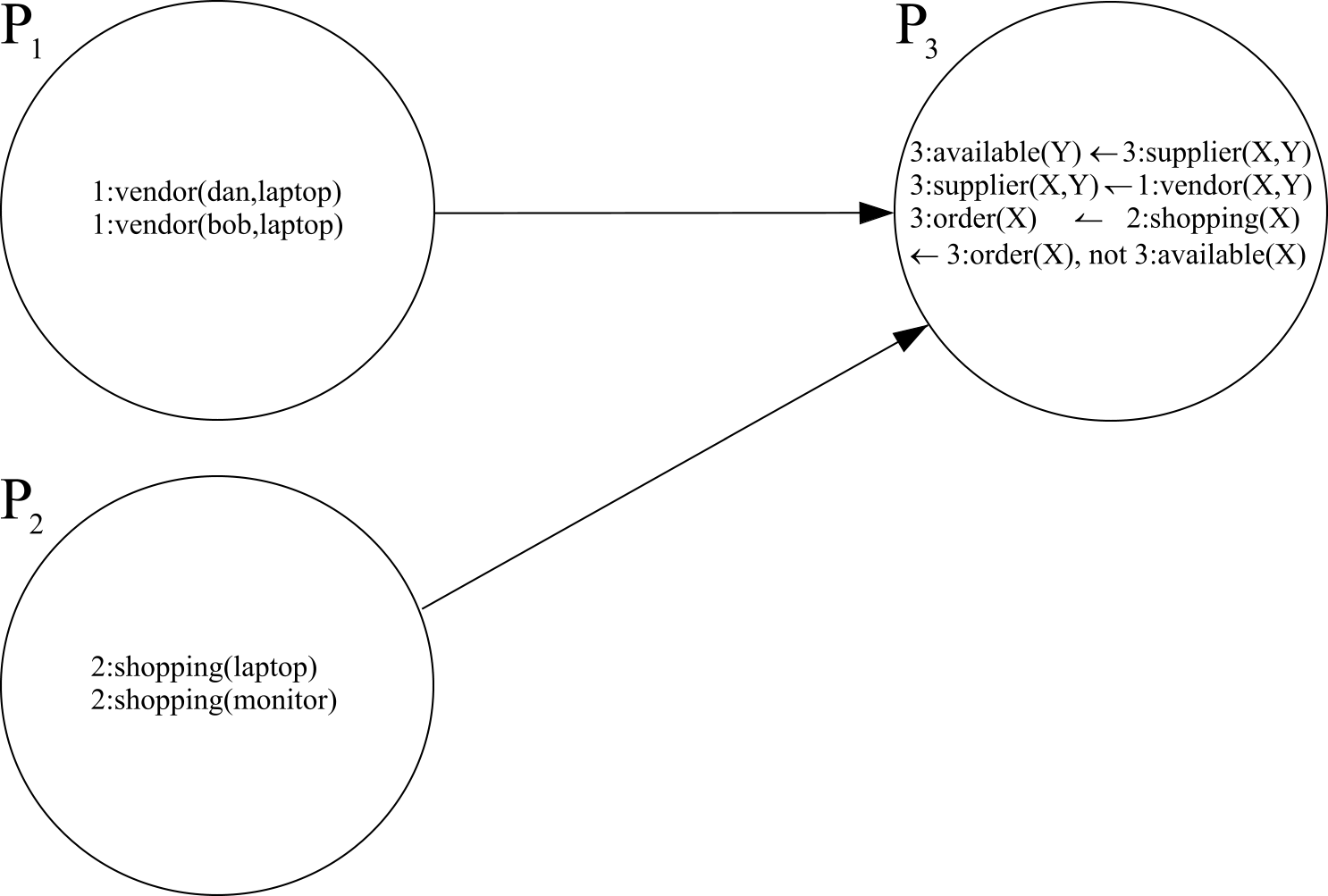}
	\caption{A P2P System with maximal and minimal mapping rules}
	\label{fig-Motivating-Example-Max-Min}
\end{figure}
Consider the P2P system depicted in Figure \ref{fig-Motivating-Example-Max-Min}. 
\begin{itemize}
	\item
	Peer
	$\PP_1$ stores information about vendors of devices and contains the 
	facts:  \\
	$1\:vendor$ $(dan,laptop)$,
	whose meaning is \emph{`Dan is a vendor of laptops'} and \\
	$1\:vendor(bob,$ $laptop)$, whose meaning is
	\emph{`Bob is a vendor of laptops'}; \\
	\item
	Peer $\PP_2$ stores information about devices that should be
	ordered: \\ $2\:shopping(laptop)$ and \\ $2\:shopping(monitor)$; \\
	\item
	Peer
	$\PP_3$ contains:
	\begin{itemize}
	\item
	the integrity constraint \\
	$\leftarrow 3\:order(X),\ not \ 3\:available(X)$
	disallowing to import the order of a device that cannot be provided
	by any supplier.	
		\item
	the standard rule \\$3\:available(Y)\leftarrow 3\:supplier(X,Y)$, stating that a device $Y$ is
	available if there is a supplier $X$ of $Y$
	\item 
	  two mapping rules \\
	- $3\:order(X) \leftharpoonup 2\:shopping(X)$ that, intuitively, allows to import as many orders as possible 
	from $\PP_2$  and \\
	- $3\:supplier(X,Y) \leftharpoondown
	1\:vendor(X,Y)$ that allows to import minimal sets of supplier from $\PP_1$ able to
	provide the ordered devices. 
	\end{itemize}
	\nop{
	intuitively the first one allows to import as many orders as possible
	from $\PP_2$ ($3\:order(X) \leftharpoonup 2\:shopping(X)$), and the second
	one allows to import minimal sets of supplier from $\PP_1$ able to
	provide the ordered devices ($3\:supplier(X,Y) \leftharpoondown
	1\:vendor(X,Y)$). Moreover, in $\PP_1$ there is an integrity constraint
	disallowing to import the order of an device that cannot be provided
	by any supplier ($\leftarrow 3\:order(X),\ not \ 3\:available(X)$).
}
\end{itemize}
The intuitive meaning of the P2P system is the following: the fact
$2\:shopping(laptop)$, belonging to the adding resource $\PP_2$, can be
used to derive $3\:order(laptop)$. This fact does not violate the
integrity constraint in $\PP_3$ thanks to the repair  resource $\PP_1$ whose
role is to try to guarantee the consistency of $\PP_3$. 
In more detail, either
$1\:vendor(dan,laptop)$ or $1\:vendor(bob, laptop)$ can be
used to derive the fact $3\:available(laptop)$ and thus to satisfy
the constraint.
Therefore, the preferred scenarios of the system, called max-min weak models, contain besides the base predicates, either the facts
$\{3\:supplier(dan,$ $laptop),\ 3\:order(laptop),\ 3\:available(laptop)\}$ or the 
facts $\{3\:supplier$ $(bob,laptop),\ 3\:order$ $(laptop),\ 3\:available(laptop)\}$.\\
Observe that, we cannot act in a similar way with respect to the fact
$2\:shopping(monitor)$ belonging to $\PP_2$: no repair mechanism can be
activated in order to support the derived predicate
$3\:order(monitor)$.\\
Summarizing, the presence of the \textit{repair resource} $\PP_1$ allows the
system to fix the knowledge imported from  $\PP_2$.
~\hfill~$\Box$
\end{example}
}

In the previous example,  peer $\PP_3$ aims at
enriching its knowledge by importing from $\PP_2$ as much
knowledge as possible, and  uses $\PP_1$ to (eventually)
restore inconsistencies.  Therefore, with respect to
$\PP_3$, peer $\PP_2$ acts as an \emph{adding resource}, whereas peer $\PP_1$ acts as a \emph{repair resource}.\\
\ch{
	\nop{
	As previously pointed out there have been several proposals
	which consider the integration of information and the computation of
	queries in an open ended network of distributed peers
	\cite{Bern-etal02,BertossiB04,Calv*04,DBLP:conf/dbisp2p/CalvaneseDGLR03,Franconi*04a,DBLP:conf/pods/AbiteboulD98,TatHal04,GribbleHIRS01,DBLP:journals/tcs/FaginKMP05}.
	}
	Different alternative semantics for P2P systems, that will be extensively discussed in Section \ref{Discussion}, have been proposed in the literature. In any case, in each of them mapping rules are used as a vehicle to import data. 
	\nop{The framework, here proposed, thanks to the introduction of two different forms of mapping rules, 
	allows to characterize each peer in the neighborhood as a resource
	used either to enrich (integrate) or to fix (repair) the knowledge,
	so as to define a kind of \emph{integrate\-repair} strategy for
	each peer in the P2P setting.
}
}

\ch{	Our approach, as well as in general P2P data management systems, can be viewed as a special case of Multi-Context Systems (MCS) 
	as it models autonomous logic-based entities (peers) that interchange pieces of information
	using mapping rules. In any case, the essential feature of P2P system is that each peer may leave and join the system arbitrarily. Due to this specific dynamic nature,  the focus in the P2P context is not that of finding the explanations of inconsistencies, but just cope with them. In our work, due to the two different forms of mapping rules, each peer is given the possibility to decide how to interact with a neighbor peer: as a source used to maximize its own knowledge or as a source used to fix its own knowledge. This specific notion has not a counterpart in any of the previous works in the literature, neither in the field of MCS, nor in the field of P2P systems.}

\paragraph{\bf{Contributions.}}
The paper presents a logic-based framework for modeling the
interaction among peers. It is assumed that each peer consists of a
database, a set of standard logic rules, a set of mapping rules of
two possible types and a set of integrity constraints. In such a
context, a query can be posed to any peer in the system and the
answer is provided by using locally stored data and all the
information that can be consistently imported from the neighborhood.

 In
synthesis, the main contributions of the paper are:
\begin{itemize}
\item
\ch{The introduction of two different forms of mapping rules:
	\emph{maximal mapping rules} used to import maximal sets of
	atoms  while
		preventing inconsistency anomalies  and 	\emph{minimal mapping rules} used to fix the knowledge by importing minimal sets of atoms allowing to restore consistency. In other words, maximal mapping rules state that it is preferable to import as long as no local inconsistencies arise; whereas minimal mapping rules state that it is preferable not to import unless a local inconsistency exists.
		By using, these two forms of mapping rules a generic peer is able
 to decide
how to interact with a neighbor peer: as a source used to maximize its own knowledge or as a source used to fix its own knowledge. 
}
\\
\item
The proposal of the \emph{Max Weak Model Semantics}, in which mapping rules are used to import  \emph{as much knowledge as possible} from its neighborhood without violating local integrity constraints. \vspace{2mm}
\item
The proposal of the \emph{Min Weak Model Semantics}, in which the  P2P system can be locally inconsistent and the information provided by the neighbors is  used to restore
consistency, that is to only integrate the missing portion of a correct,
but incomplete database.\vspace{2mm}
\item
The \textit{Max-Min Weak Model Semantics} that
unifies the previous two different
perspectives captured by the Max and Min Weak Model Semantics. 
This more general declarative  semantics,
allows to characterize each peer in the neighborhood as a resource
used either to enrich (integrate) or to fix (repair) the knowledge,
so as to define a kind of \emph{integrate\-repair} strategy for
each peer in the P2P setting. 
\nop{More specifically it: i) uses the maximal
mapping rules among peers to import maximal sets of atoms; ii) uses
the minimal mapping rules among peers to import only minimal sets of
atoms; iii) ensures local consistency for each peer. \vspace{2mm}
}
\item
The definition of an alternative  characterization of the Max-Min Weak Model Semantics (resp. Max Weak Model Semantics and Min Weak Model Semantics) obtained by rewriting mapping rules into prioritized rules. Therefore, a P2P system $\PS$ is rewritten into an equivalent prioritized logic program, $Rew(\PS)$, 
such that the max-min weak models of $\PS$ (resp. maximal weak models and minimal weak models) are 
the preferred stable models of $Rew(\PS)$.\vspace{2mm}
\item
\chg{
Results on the complexity of answering queries. The paper,  by considering analogous results on stable model
semantics for prioritized logic programs, proves that for disjunction-free
($\vee-free$) prioritized programs
deciding whether an interpretation 
$M$ is a max-min weak model (resp. maximal weak models and minimal weak model) of $\PS$ is 
$co\NP$ complete; 
deciding whether
an atom is \emph{true} in some max-min weak model (resp. maximal weak models and minimal weak model) is
$\Sigma_2^p$-complete, whereas deciding whether an atom is
\emph{true} in every preferred model is $\Pi_2^p$-complete
\cite{SakIno00}. Moreover, the paper also provides results on the existence of a max-min weak model (resp. maximal weak model and minimal weak model) showing that the problem 
is in $\Sigma_2^p$.}\vspace{2mm} 
\item
An extensive section,  Discussion, reporting different features of the proposal.
In more detail, the practical aspects of the proposal are highlighted 
and several additional and alternative issues, arising from the basic framework, are presented: a technique allowing to deal with P2P systems locally inconsistent; a  deterministic semantics, derived from the max weak model semantics, allowing to assign a unique three value model to particular types of P2P systems;  
a  polynomial distributed algorithm  for its computation;
a system prototype.

\end{itemize}

\paragraph{\bf{Structure of the paper.}}
\ch{
The remainder of the paper is organized as follows. Section \ref{background} introduces relevant background information. Section \ref{P2PSystem} describes the  syntax of  P2P
systems. Section \ref{sem} describes alternative semantics, namely the \emph{Max Weak Model Semantics} in
\cite{CarGre*06a},  the \emph{Min Weak Model Semantics} in
\cite{CarZum12} and introduces a new formal declarative semantics, called \emph{Max-Min Weak Model Semantics},
that unifies the previous two into a more general perspective. Moreover, it introduces, for each of the proposed semantics, an alternative 
characterization, modeled in terms of logic
programs with priorities. Section \ref{compl} presents results on
computational complexity, Section \ref{Discussion} focuses on some relevant discussions related to practical aspects of the proposed framework and  Section \ref{RW} introduces a comprehensive discussion of related works. Finally, Section \ref{Concl}
reports concluding remarks and directions for further research.
}

\section{Background}
\label{background}
We assume that there are finite sets of \textit{predicates}, constants
and variables \cite{DBLP:books/aw/AbiteboulHV95}. A \emph{term} is either a constant or a variable. An
\emph{atom} is of the form $p(t_1,\dots,t_n)$ where $p$ is a
predicate and $t_1,\dots,t_n$ are terms. A \emph{literal} is
either an atom $A$ or its negation $not\ A$.
\ch{
As in this work we use the \textit{Closed World Assumption}, we adopt \textit{negation as failure}.
}
\ch{
A \emph{rule} is of the
form:
\begin{itemize}
	\item $H \leftarrow \B$,
	where $H$ is an atom and $\B$ is a conjunction of literals or
	\item $\leftarrow \B$,
	where $\B$ is a conjunction of literals. 
\end{itemize} 
$H$ is called {\em head} of the
rule and $\B$ is called {\em body} of the
rule. A rule of the form $\leftarrow \B$ is also called 
\emph{constraint}.
}
A program $\PP$ is a finite set of rules. $\PP$ is said to be
positive if it is negation free. 
 
\ch{
An \textit{exclusive disjunctive rule} is the form
$A \oplus A' \leftarrow \B$ and it is a notational shorthand 
for $A \leftarrow \B \wedge not\ A'$, $A' \leftarrow \B\wedge not\ A$
and $\leftarrow A\wedge A'$\footnote{We use for the operator \textit{and} both '$,$' and '$\wedge$'. }. Its intuitive meaning is that if $\B$ is
\emph{true} then exactly one of $A$ and $A'$ must be \emph{true}.
}

It is assumed that programs are \emph{safe}, i.e.
variables occurring in the head or in negated body literals are
range restricted as they occur in some positive body literal. 

An atom (resp. literal, rule, program) is \textit{ground} if no variable occurs in it.
A ground atom is also called \textit{fact}.
\ch{
\chn{The set of ground instances of an atom $a$} (resp. literal $l$, rule $r$, program $\PP$), denoted by $ground(a)$ (resp. $ground(l)$, $ground(r)$, $ground(\PP)$) is
built by replacing variables with constants in all possible ways. An
interpretation is a set of facts.
} The truth value of ground
atoms, literals and rules with respect to an interpretation $M$ is
as follows: $val_M(A) = (A \in M)$, $val_M(not\ A) = not\ val_M(A)$,
$val_M(L_1,\dots,L_n) = min\{val_M(L_1), \dots, val_M(L_n)\}$ and \
$val_M(A\leftarrow L_1,\dots,$ $L_n) $ $ = val_M(A)\geq
val_M(L_1,$ $\dots,L_n)$, where $A$ is an atom, $L_1,\dots, $ $ L_n$
are literals and $true > false$. An interpretation $M$ is a model
for a program $\PP$, if all rules in $ground(\PP)$ are \emph{true}
w.r.t. $M$. A model $M$ of a program $P$ is said to be minimal if there is no model
$N$ of $P$ such that $N \subset M$. We denote the set of minimal models of
a program $\PP$ with $\MM(\PP)$. Given an interpretation $M$ and a
predicate $g$, $M[g]$ denotes the set of $g$-tuples in $M$.
The semantics of a positive program $\PP$ is given by its unique
minimal model which can be computed by applying the {\em immediate
consequence operator} ${\bf T}_{\PP}$ until the fixpoint is reached
($\,{\bf T}_{\PP}^\infty ( \emptyset)\,$). The semantics of a
program with negation $\PP$ is given by the set of its stable
models, denoted as  $\SM(\PP)$. An interpretation $M$ is a
\emph{stable model} of $\PP$ if $M$ is the
unique minimal model of the positive program $\PP^M$, where $\PP^M$
is obtained from $ground(\PP)$ by: (i) removing all rules $r$ such
that there exists a negative literal $not\ A$ in the body of $r$ and
$A$ is in $M$ and (ii) removing all negative literals from the
remaining rules \cite{GelLif88}. It is well known that stable models
are minimal models (i.e. $\SM(\PP) \subseteq \MM(\PP)$).
\nop{ and that for
negation free programs, minimal and stable model semantics coincide
(i.e. $\SM(\PP) = \MM(\PP)$)}.

\subsection{Prioritized Logic Programs}

Several works have investigated various forms of
priorities into logic languages
\cite{BreEit99,Bre-etal*03,Del*03,SakIno00}.
In this paper we refer to the extension proposed in \cite{SakIno00}.
\\
\chrr{
A \emph{preference relation} $\succeq$ among ground atoms is
defined as follows. 
For any ground atoms $e_1$ and $e_2$, if $e_1 \succeq e_2$ then we say that
$e_1$  \emph{has a higher priority than} $e_2$.
$e_1 \succ e_2$ stands for $e_1 \succeq e_2$ and $e_2 \not\succeq e_1$.
The statement $e_1 \succeq e_2$ is called a \emph{priority}.
The statement $p_1(x) \succeq p_2(y)$, where $x$ and $y$ are tuples containing variables, stands
for every priority $p_1(s) \succeq p_2(t)$, where $s$ and $t$ are instances of $x$ and $y$ respectively.
\\
If $p(x) \succ p(y)$, $p(x)$ and $p(y)$ do not have common ground instances.
Indeed, assuming that there is a ground atom $p(s)$ which
is an instance of $p(x)$ and $p(y)$, the statements $p(s) \succeq p(s)$ and $p(s) \not\succeq p(s)$
would hold at the same time (a contradiction).\\
Given a set $\Phi$ of priorities, we define the closure $\Phi^*$ as the set of priorities which are reflexively or transitively derived using priorities in $\Phi$.
\\
\nop{
The relation $\succeq$ is transitive
and reflexive.\\
Let $\Phi$ be a set of priorities,  $\Phi^*$ its ground version and $a_1$ and $a_2$ ground atoms. We write: 
\begin{itemize}
	\item $\Phi\models (a_1 \succeq a_2)$, if 
	$(a_1 \succeq a_2) \in \Phi^*$;
	\item $\Phi\models (a_1 \succ a_2)$, if 
	$(a_1 \succeq a_2) \in \Phi^*$ and $(a_2 \succeq a_1) \not\in \Phi^*$.
\end{itemize}
}
Let $\M$ be a class of sets of ground atoms and $\Phi$ a set of priorities.
The relation $\sqsupseteq$ is defined over the sets of $\M$
as follows. For any sets $M_1, M_2$ and $M_3$ of $\M$:
\begin{itemize}
\item
$M_1 \sqsupseteq M_1$; 
\item $M_1 \sqsupseteq M_2$ if
$\exists \ e_1 \in M_1 \setminus M_2, \ \exists \ e_2 \in M_2 \setminus M_1$ such that
$(e_1 \succeq e_2) \in \Phi^*$ and
$\not\exists \ e_3 \in M_2 \setminus M_1$ such that $(e_3 \succ e_1)\in \Phi^*$; 
\item 
if $M_1
\sqsupseteq M_2$ and $M_2 \sqsupseteq M_3$, then $M_1 \sqsupseteq
M_3$.	
\end{itemize}
\noindent
If $M_1 \sqsupseteq M_2$ holds, then we say that $M_1$ is \emph{preferable} to
$M_2$ w.r.t. $\Phi$. Moreover, we write $M_1 \sqsupset M_2$ if $M_1 \sqsupseteq
M_2$ and $M_2 \not\sqsupseteq M_1$.
A set $M$ is a \emph{preferred} set of $(\M,\Phi)$
if $M$ is in $\M$ and there is no set $N$ in $\M$
such that $N \sqsupset M$.
The class of preferred sets of
$(\M,\Phi)$ will be denoted by $\PS(\M,\Phi)$.
}
A \emph{prioritized logic program} (PLP) is of the form
$(\PP,\Phi_1,\ldots,$ $\Phi_n)$ where $\PP$  is a logic program and
$\Phi_1,\ldots,\Phi_n$, with $n\geq 1$, are sets of priorities. The
\emph{preferred stable models} of $(\PP,\Phi_1,\ldots,\Phi_n)$ denoted as
$\PSM(\PP,\Phi_1,\ldots,\Phi_n)$ are the stable models of $\PP$
selected by applying consecutively the sets of priorities
$\Phi_1,\ldots,\Phi_n$. More formally:

\noindent
\begin{itemize}
\item
$\PSM(\PP,\Phi_1)=\PS(\SM(\PP),\Phi_1)$
\item
$\PSM(\PP,\Phi_1,\ldots,\Phi_n)=\PS(\PSM(\PP,\Phi_1,\ldots,\Phi_{n-1}),\Phi_n)$
\end{itemize}
\normalsize

\section{P2P Systems}\label{P2PSystem}

\ch{
A peer identifier is a number $i\in \mathbb{N}^+$.
A \emph{(peer) predicate} is a pair $i\:p$, where $i$ is a
peer identifier and $p$ is a predicate\footnote{\ch{Whenever the reference to a \textit{peer predicate} (resp. \textit{peer atom, peer literal, peer fact, peer rule, peer standard rule, peer integrity constraint, peer maximal mapping rule, peer minimal mapping rule}) is clear from the context, the term \textit{peer} can be omitted.}}.
A \emph{(peer) atom} $A$ is of the form $i\:p(X)$, where $i$ is a
peer identifier, $p(X)$ is an atom and $X$ is a list of terms. A \emph{(peer) literal} is a peer atom $A$ or its negation $not\ A$.
A conjunction $\B=i\:p_1(X_1),\dots,i\:p_m(X_m),$ $not\ i\:p_{m+1}(X_{m+1}),\dots,not\ i\:p_n(X_n), \varphi$, 
where $\varphi$ is a conjunction of built-in atoms\footnote{\ch{A \emph{built-in atom} is
of the form $\theta(X,Y)$, where $X$ and $Y$ are terms and $\theta\in\{<,>,\leq,\geq,=,\neq\}$. It is also denoted as $X\ \theta\ Y$.}}, will be also denoted as $i\:(p_1(X_1),\dots,p_m(X_m),$ $not\ p_{m+1}(X_{m+1}),\dots,not\ p_n(X_n),\varphi)$.
}

\ch{
\begin{definition} \textsc{[Peer Rule].}\label{PeerDef}
A \emph{(Peer) rule} can be of one of the following types:
\begin{enumerate}
	\item
	\chn{
	\textit{(Peer) standard rule.}\\ 
	It is of the form $H \leftarrow \B$, where $H=i\:h(X)$ and $\B=j\:(p_1(X_1),\dots,p_m(X_m),$ $not\ $ $p_{m+1}(X_{m+1}),\dots,not$ $ p_n(X_n),\varphi)$.
	}
	\item
	\textit{(Peer) integrity constraint.}\\
	It is of the form $\leftarrow \B$, where
	$\B=i\:(p_1(X_1),\dots,p_m(X_m),$ $not\ p_{m+1}(X_{m+1}),\dots,not$ $ p_n(X_n),\varphi  )$,
	\item
	\textit{(Peer) maximal mapping rule.}\\
	It is of the form $H \leftharpoonup \B$, where $H=i\:h(X)$, $\B=j\:(p_1(X_1),\dots,p_m(X_m),\varphi)$ and $i \neq j$.
	\item
	\textit{(Peer) minimal mapping rule.}\\
	It is of the form $H \leftharpoondown \B$, where $H=i\:h(X)$, $\B=j\:(p_1(X_1),\dots,p_m(X_m),\varphi)$ and $i \neq j$.~\hfill~$\Box$
\end{enumerate}
\end{definition}
}

\ch{
\noindent
In the previous definition, $i$ (resp. $j$) is the \textit{peer identifier} (resp. \textit{source peer identifier}) of the rule, $H$ is the \emph{head} of the rule and $\B$ is the \emph{body} of rule. With the term \textit{mapping rule} we refer to
a maximal mapping rule or to a minimal mapping rule.
The
concepts of \emph{ground rule}, \emph{fact} \chn{and \textit{interpretation}} are similar to
those reported in Section \ref{background}. Given a fact $i\:p(x)$, $i$ is its peer identifier.
}

\nop{
\ch{
The definition of a predicate $i\:p$ consists of the set of rules in whose head $i\:p$ occurs and the set of ground atoms whose predicate is $i\:p$.
} 
}
\ch{
In our setting, a predicate is of exactly one of the following three types: \emph{base predicate}, \emph{derived predicate}
and \emph{mapping predicate}. 
\chrr{A derived predicate is a predicate occurring in the head of a standard rule, a mapping predicate is a predicate occurring in the head of a mapping rule. If a predicate is neither a derived predicate nor a mapping predicate, then it is a base predicate.}
\nop{A base predicate is defined by a set
of ground atoms; a derived predicate is defined by a set of
standard rules and a mapping predicate is defined by a set of
mapping rules.}
\\
\noindent
An atom $i\:p(X)$ is a \emph{base atom} (resp. \emph{derived atom}, \emph{mapping atom})
if $i\:p$ is a base predicate (resp. derived predicate, mapping predicate).}

\chn{
The intuitive meaning of a standard rule is that \textit{whenever its body
is true, its head has to be true}.
This meas that an interpretation $M$ satisfies a standard rule $r$ if
for each ground instance $r'$ of it, $M$ does not satisfy the body of
$r'$ or $M$ satisfies the head of $r'$.
\\
The intuitive meaning of an integrity constraint is that \textit{its body has to be false}.
Therefore, an interpretation $M$ satisfies an integrity constraint  $c$ if
for each ground instance $c'$ of it, $M$ does not satisfy the body of
$c'$.\\
The intuitive meaning of a maximal rule is that \textit{whenever its body
is true, its head has to be true if it does not violate (directly or indirectly) any integrity constraint}.\\
Finally, the intuitive meaning of a minimal rule is that \textit{whenever its body
is true, its head has to be true if it prevents the violation (directly or indirectly) of some integrity constraint}.\\
In the following sections, we will see how the semantics of a maximal mapping rule and a minimal mapping rule can be captured 
by an\textit{ exclusive disjunctive rule} and a \textit{priority}.
}

Given an interpretation $M$, $M[\D]$ (resp. $M[\LP]$, $M[\MP]$) denotes the subset
of base atoms (resp. derived atoms, mapping atoms) in $M$.

\ch{
\begin{definition} \textsc{[P2P SYSTEM].}
	A \emph{peer} $\PP_i$, with a peer identifier $i$,
	is a tuple $\<\D_i, \LP_i, \MP_i, \IC_i\>$, where 
\begin{itemize} 
	\item
	$\D_i$ is a set of
	facts whose peer identifier is equal to $i$ (\emph{local database}); 
	\item
	\chn{
	$\LP_i$ is a set of standard rules whose peer identifier and source peer identifier are equal to $i$;
	\vspace{-5mm}
	} 
	\item
	$\MP_i$ is a
	set of mapping rules whose peer identifier is equal to $i$ and 
	\item
	$\IC_i$ is a set of constraints over predicates defined by $\D_i$, $\LP_i$ and $\MP_i$ whose peer identifier is equal to $i$.
\end{itemize}
\noindent A \emph{P2P system} $\PS$ is a set of peers $\{\PP_1,\dots,\PP_n\}$ s.t. for each source peer identifier $j$ occurring in its mapping rules, $j\in[1..n]$.~\hfill~$\Box$
\end{definition}
}

Given a peer $\PP_i=\<\D_i, \LP_i, \MP_i, \IC_i\>$, we denote as:
\begin{itemize} 
\item
$\overline{\MP_i}$ 
the subset of maximal mapping rules in $\MP_i$ 

\item 
$\underline{\MP_i}$ 
the subset of minimal mapping rules in $\MP_i$.
\end{itemize} 
\noindent
Clearly, 	$\MP_i = \overline{\MP_i} \cup \underline{\MP_i}$.
Without loss of generality, we assume that every mapping predicate is defined
by only one mapping rule of the form $i\:p(X) \leftharpoondown j\:q(X)$ (resp. $i\:p(X) \leftharpoonup j\:q(X)$). 
\ch{
Indeed, a mapping predicate $i\:p$ consisting of $n$
rules of the form $i\:p(X) \Leftarrow_k j_k\:\B_k$, with $\Leftarrow_k \in \{\leftharpoonup,\leftharpoondown\}$ and $k \in [1..n]$,
can be rewritten into $2\cdot n$ rules of the form 
$i\:p_k(X) \Leftarrow_k j_k\:\B_k$ and 
$i\:p(X) \leftarrow i\:p_k(X)$ with $k \in [1..n]$.
 \chrr{Observe that, $i\:p$ becomes a derived predicates and $i\:p_k(X) $, with $k \in [1..n]$, are new mapping predicates}. 
Moreover, there is no loss of generality 
in considering mapping rules having a positive body.
Indeed, allowing negation in the body of mapping rules,
a mapping rule $H \Leftarrow \B(X)$, where $\Leftarrow \in \{\leftharpoonup,\leftharpoondown\}$ and 
$\B(X)=j\:(p_1(X_1),\dots,p_m(X_m),$ $not\ $ $p_{m+1}(X_{m+1}),\dots,not$ $ p_n(X_n),\varphi)$,
could be rewritten into the mapping rule 
$H \Leftarrow j\:c(X)$ and the standard rule $j\:c(X)\leftarrow \B(X)$.
}

\vspace{2mm}
\noindent
Given a P2P system $\PS = \{\PP_1,\dots,\PP_n \}$, where
$\PP_i = \<\D_i, \LP_i, \MP_i, \IC_i \>$  with $i \in [1..n]$, the sets $\D, $ $ \LP, $ $ \MP$, $\IC$, $\overline{\MP}$ and $\underline{\MP}$ denote,
respectively, the global sets of ground facts, standard rules, mapping rules, integrity constraints, maximal mapping rules and minimal mapping rules that is:
\begin{itemize} 
\item $\D = \bigcup_{i \in [1..n]}\D_i$, 
\item $\LP = \bigcup_{i \in [1..n]}\LP_i$,
\item $\MP = \bigcup_{i \in [1..n]}\MP_i$, 
\item $\IC = \bigcup_{i \in [1..n]}\IC_i$,
\item $\overline{\MP} = \bigcup_{i \in [1..n]}\overline{\MP_i}$,
\item $\underline{\MP} = \bigcup_{i \in [1..n]}\underline{\MP_i}$ 
\end{itemize} 

In the rest of the  paper, with a little abuse of notation, $\PS$ will be denoted both with the 
tuple $\<\D, \LP, \MP, \IC\>$ and with the set
$\D \cup \LP \cup \MP \cup \IC$. 
\\
\ch{Moreover, we call a P2P system only containing maximal mapping rules, a \textit{maximal P2P system} and   a P2P system only containing minimal mapping rules, a \textit{minimal P2P system}.}

\ch{
A peer
$\PP_i = \< \D_i, \LP_i, \MP_i,$ $\IC_i \>$ is \emph{locally consistent} if $\SM(\D_i \cup \LP_i\cup \IC_i)\neq \emptyset$. 
A P2P system whose peers are locally consistent is \textit{locally consistent}. A peer (resp. P2P system) that is not locally consistent is \textit{locally inconsistent}.
}

Given a mapping rule 
$r=H \leftharpoonup \B$ (resp. $r=H \leftharpoondown \B$),
the corresponding standard logic rule $H \leftarrow \B$ will be denoted as  $St(r)$.

\noindent
Analogously, given a set of mapping rules $\MP$, $St(\!\MP)\!= \{
St(r)\ |\ r \in \MP \}$ and given a P2P system $\PS = \D \cup \LP
\cup \MP \cup \IC$, $St(\PS) = \D \cup \LP \cup St(\MP) \cup
\IC$.

\ch{
	In this context an interpretation is a set of peer facts.
The truth value of a peer fact (resp. literal, rule, maximal mapping rule, minimal mapping rule) with respect to an interpretation $M$ is as follows:
\begin{itemize}
\item $val_M(A)=(A\in M)$,
\item $val_M(not\ A)=not\ val_M(A)$,
\item $val_M(L_1,\dots,L_n)=min\{val_M(L_1),\dots,val_M(L_n)\}$,
\item $val_M(H \leftarrow \B)=val_M(H) \geq val_M(\B)$,
\item $val_M(H \leftharpoonup \B)=val_M(H) \leq val_M(\B)$,
\item $val_M(H \leftharpoondown \B)=val_M(H) \leq val_M(\B)$.
\end{itemize}
}
\ch{
Therefore, while a standard rule is satisfied if its body is \textit{false} or  its  body is \textit{true} and its head is
\textit{true}, a mapping rule is satisfied if its body is \textit{true} or its body is \emph{false} and its head is \textit{false}.
}
\nop{-----------------------FOL SEmantics
\vspace{2mm}
\paragraph{\bf FOL semantics.}
The FOL semantics of a
P2P system $\PS = \{ \PP_1,\dots,$ $\PP_n \}$
is given by the minimal model semantics of $\PS = \D \cup \LP \cup \MP \cup \IC$.
For a given P2P system $\PS$,
$\MM(\PS)$ denotes the set of minimal models of $\PS$.
As $\D \cup \LP \cup \MP$ is a positive program, $\PS$ may
admit zero or one minimal model.
In particular, if $\MM(\D \cup \LP \cup \MP) = \{ M \}$ then
$\MM(\PS) = \{ M \}$ if $M \models \IC$, otherwise
$\MM(\PS) = \emptyset$.
The problem with such a semantics is that local inconsistencies
make the global system inconsistent.

-----------------------------------------}

\section{Semantics for P2P Systems}\label{sem}
\ch{
Recent literature proposed different semantics for P2P systems that will be discussed in Section \ref{RW}.
The simplest semantics for a P2P system is the First Order Logic (FOL) semantics obtained by interpreting mapping rules as standard rules. The FOL semantics of a
P2P system $\PS = \< \D, \LP, \MP, \IC\>$ is given by the set
of minimal models of $(\D \cup \LP \cup St(\MP)\cup \IC)$.
The problem with the FOL semantics is that it leads to  global inconsistency ($MM(\D \cup \LP \cup St(\MP)\cup \IC)=\emptyset$) as soon as an atom imported in a peer
causes a violation of one of its integrity constraints.
\\
It's clear that \textit{more robust semantics}, derived by assuming \textit{more flexible behaviors} of mapping rules, are needed.\\
\chn{
It is worth noting that, in the FOL semantics, classical negation is used. In this paper, instead, we adopt negation as failure suited for all the non  monotonic semantics here presented.
}
\\
Our insight is that a peer of a P2P system can use its mapping rules to import from its neighborhood either \textit{as much knowledge as possible preserving its consistency} or just \textit{the knowledge that is strictly needed to restore the consistency of the system}. \\
Starting from this idea, in this section we first present two alternative semantics for P2P systems:
the \textit{Max Weak Model Semantics} and the \textit{Min Weak Model Semantics}.\\
In the Max Weak Model Semantics, the peers of a P2P system only have maximal mapping rules and use them to import maximal sets of facts not violating local integrity constraints. \\ 
In the Min Weak Model Semantics, the peers of a P2P system only have minimal mapping rules and use them to import minimal sets of facts that are strictly needed to restore the consistency of the system.\\ 
In the Max-Min Weak Model Semantics, the peers of a P2P system have maximal and minimal mapping rules and
unifies the two previous perspectives.
A peer can use each neighbor as a resource either to enrich (integrate) or to fix (repair) its knowledge,
adopting a kind of \emph{integrate-repair} strategy. \\ 
All these semantics guarantee that a P2P system that is 
locally consistent admits at least a model, i.e. remains consistent.
}


In order to present the three different semantics first of all we introduce the concept of weak model, that is common to all of them.
\ch{
\begin{definition} \label{definition-weak-model}
\textsc{[Weak Model].} \ \em
Given a P2P system $\PS = \D \cup \LP \cup \MP \cup \IC$,
an interpretation $M$ is a \emph{weak model} for $\PS$
if $\{M\} = \MM(St(\PS^M))$, where $\PS^M$ is the program
obtained from $ground(\PS)$ by:
\begin{itemize}
	\item removing all peer rules $r$ such that a negative literal $not\  A$ occurs in the body of $r$ and $A$ is not in $M$;
	\item removing from the remaining peer rules each negative literal.
	\item removing all mapping rules whose head is not in $M$;
\end{itemize}
The set of weak models of $\PS$  will be denoted as
$\WM(\PS)$.
~\hfill $\Box$
\end{definition}
}

\ch{
Observe that, $St(\PS^M)$ is an Horn program and it 
can be partitioned into a set of standard rules $\Pi$, a set of integrity constraints $\Sigma$ and a set of facts $D$ (i.e. $St(\PS^M)=\Pi \cup \Sigma \cup D$).
\\
As $\Pi\cup D$ is a positive normal program, it admits exactly one minimal model $N$. Therefore, $N$ is the minimal model of $St(\PS^M)$ if
$N\models \Sigma$, otherwise $St(\PS^M)$ does not admit any minimal model. If $N=M$ then $M$ is a weak model of $\PS$.\\
Note that,
the definition of weak model presents interesting analogies with the definition of stable model (see Section \ref{background}). \\
Indeed, given a logic program $P$, an interpretation $M$ is a stable model of $P$ if $M$ is the minimal model of the reduct $P^M$, where the reduct is obtained by removing from $ground(P)$ each rule $r$ such that a negative literal $not\ A$ occurs in the body of $r$ and $A$ is not in $M$ and removing from the remaining rules each negative literal.
The fact that $M$ is a minimal model of the reduct $P^M$ ensures that each atom $H \in M$ is \textit{supported}, i.e. there is a rule in $ground(P)$ whose head is $H$ and whose body is satisfied by $M$.\\
Similarly, in Definition \ref{definition-weak-model} the fact 
that $M$ is a minimal model of $St(\PS^M)$ ensures that each atom is supported. In particular, for each mapping atom $H\in M$ there is a mapping rule in $ground(\MP)$ whose head is $H$ and whose body is satisfied by $M$.
}

\ch{
\begin{example}\label{Ex3-maximal}
\begin{figure}[h]
	\centering
	\includegraphics[width=0.8\textwidth]{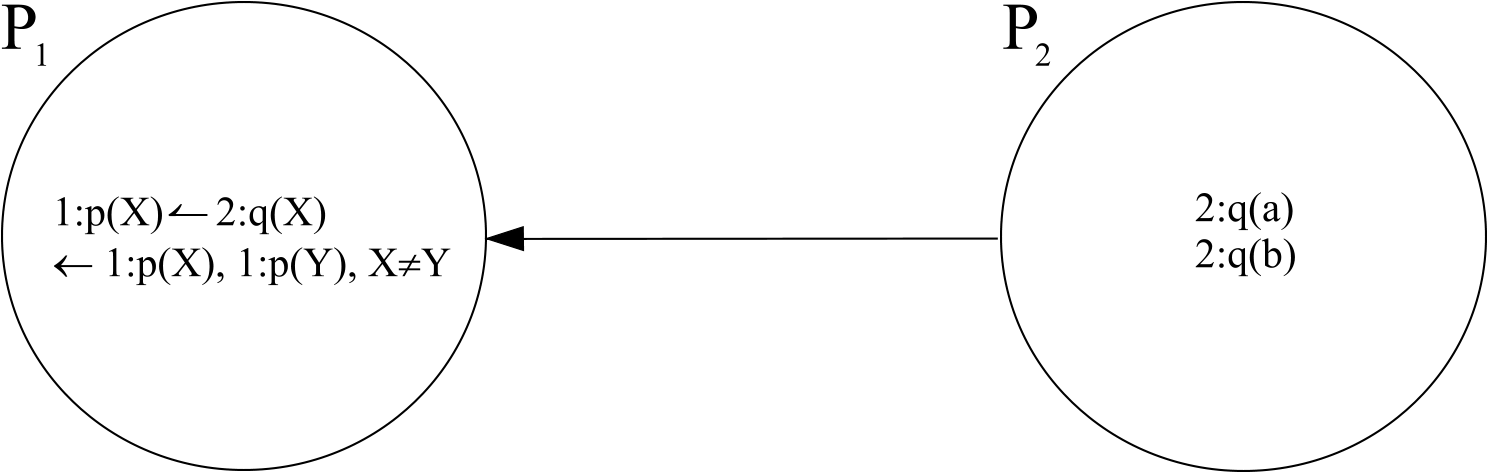}
	\caption{A System $\PS$}
	\label{fig3-maximal}
\end{figure}
\noindent Consider the P2P system $\PS$ depicted in Figure
\ref{fig3-maximal}. $\PP_2$ contains the facts $2\:q(a)$ and $2\:q(b)$, whereas
$\PP_1$ contains
the maximal mapping rule $1\:p(X) \leftharpoonup 2\:q(X)$ and
the  constraint $\leftarrow 1\:p(X), $ $1\:p(Y), X\!\neq\!Y$.
The weak models of the system are $M_1=\{2\:q(a),2\:q(b)\}$,
$M_2=\{2\:q(a),2\:q(b),$ $1\:p(a)\}$ and $M_3=\{2\:q(a),2\:q(b),1\:p(b)\}$.
~\hfill $\Box$
\end{example}
}

We shall denote with $M[\D]$ (resp. $M[\LP]$, $M[\MP]$, $M[\overline{\MP}]$, $M[\underline{\MP}]$) the set
of ground atoms of $M$ which are defined in $\D$
(resp. $\LP$, $\MP$, $\overline{\MP}$, $\underline{\MP}$).

Given a pair $P=(A,B)$, where $A$ and $B$ are generic objects, $P[1]$ (resp. $P[2]$)
denotes the object $A$ (resp. $B$).

\chn{
The next proposition shows that for a  P2P system $\PS = \D \cup \LP \cup \MP \cup \IC$ having only positive rules in $\LP$, checking if an interpretation $M$ is 
a weak model is simpler because a simpler
reduct involving only $ground(MP)$ can be used.
}
\chn{
	\begin{proposition} \label{weak-model-positive-lp}
		\em
		Given a P2P system $\PS = \D \cup \LP \cup \MP \cup \IC$ s.t.
		no negation occurs in $\LP$,
		an interpretation $M$ is a \emph{weak model} for $\PS$
		if and only if $\{M\} = \MM(St(\PS_M))$, where $\PS_M$ is the program
		obtained from $ground(\PS)$ by removing all mapping rules whose head is not in $M$.
	\end{proposition}
	\textbf{Proof.}	
	$St(\PS^M)$ can be obtained from $St(\PS_M)$ by simply removing from 
	$ground(\IC)$ each negative literal $not\ A$ s.t. $A\not\in M$ that is
	if $St(\PS_M)=\Pi \cup \overline{\Sigma} \cup \D$, then $St(\PS^M)=\Pi \cup \Sigma \cup \D$, where $\Sigma$ is obtained from $\overline{\Sigma}$ by removing each negative literal $not\ A$ s.t. $A\not\in M$. \\
$(\Rightarrow)$
As $M$ is a weak model for $\PS$, it is the minimal model
of $\Pi \cup \D$ and $M\models \Sigma$. As all the negative literals $not\ A$
occurring in $\overline{\Sigma}$ are s.t. $A\not\in M$, it follows that $M\models \overline{\Sigma}$. Therefore, $M$ is a minimal model of  $St(\PS_M)$.\\
$(\Leftarrow)$
If $M$ is the minimal model of
$St(\PS_M)$,  it is the minimal model
of $\Pi \cup \D$ and $M\models \overline{\Sigma}$. 
Let us consider an integrity constraint $ic\in \Sigma$ 
(observe that, the body of $ic$ only contains positive literals).
There are the following cases:
\begin{itemize}
	\item $ic \in ground(IC)$. In this case, no negative atom has been removed from the original integrity constraint in order to obtain $ic$.
	We have that $ic\in \overline{\Sigma}$ and $M\models ic$.
	\item $ic \not\in ground(IC)$. In this case $ic$ has been obtained 
	from a ground integrity constraint $\overline{ic}\in ground(IC)$. All negative literals $not\ A$ removed from $\overline{ic}$ in order to obtain
	$ic$ are not in $M$ (otherwise $ic$ could not belong to $\Sigma$).
	Moreover, $\overline{ic}\in \overline{\Sigma}$. As $M\models \overline{ic}$ and each negative literal occurring in $\overline{ic}$ is true w.r.t. $M$, at least a positive literal occurring in $\overline{ic}$ has to be false w.r.t. $M$. It follows that $M\models ic$.
\end{itemize}
Therefore, $M\models \Sigma$ and $M$ is a minimal model of $St(\PS^M)$, that is $M$ is a weak model for $\PS$.
~\hfill $\Box$
}

\subsection{Max Weak Model Semantics}
\label{maximal}
In previous works  \cite{CarGre*06a,Car*06b,SUM2007,ISMIS2008,FOIKS2012},  the authors introduced the \emph{Max Weak Model Semantics}. 
\nop{
Under this semantics only facts not making
the peer inconsistent are imported, and the maximal
weak models are those in which peers import maximal sets of facts
not violating integrity constraints. Therefore,  each peer,  in the system can be thought as an integrate resource.  
}

We recall that a \emph{maximal mapping rule} (see Definition \ref{PeerDef}) is of the form $H  \leftharpoonup \B$.
Intuitively, $H  \leftharpoonup \B$ means that if the body
conjunction $\B$ is \emph{true} in the source peer, the atom $H$ will
be imported in the target peer (that is $H$ is \emph{true} in the
target peer) only if it does not imply (directly or indirectly) the violation of some constraints.
\ch{
In this section, we assume that all mapping rules of a P2P system $\PS=\<D,LP,MP,IC\>$ are maximal mapping rules i.e. $\PS$ is a maximal P2P system.
}
\ch{
\begin{example}
Consider a P2P system consisting of two peers $\PP_1$ and $\PP_2$, where:\\ 
$\PP_1 = \< \{1\:q(b)\} , \emptyset,  \emptyset, \emptyset, \emptyset\>$\\
$\PP_2 = \< \{2\:s(a)\},  \{2\:r(X) \leftarrow 2\:p(X);\ \  2\:r(X) \leftarrow 2\:s(X)\},$ \\
\hspace*{8mm}
$\{2\:p(X) \leftharpoonup 1\:q(X)\}, \ \{\leftarrow 2\:r(X), 2\:r(Y), X\!\neq\!Y\} \>$\\
In this case, the fact $2\:p(b)$ cannot be imported in $\PP_1$ as it
indirectly violates the integrity constraint.
\end{example}
}

\begin{definition} \label{definition-maximal-weak-model}
\textsc{[Maximal Weak Model].} \ch{ 
Given a maximal P2P system $\PS$ and two weak models $M$ and $N$ of $\PS$}, $M$ is said \emph{max-preferable}
to $N$,  and is denoted as $M \sqsupseteq_{Max} N$, if $M[\MP] \supseteq N[\MP]$.
Moreover, if $M \sqsupseteq_{Max} N$ and $N \not\sqsupseteq_{Max} M$, then $M
\sqsupset_{Max} N$.
A weak model $M$ \ch{of $\PS$} is \emph{maximal} if there is no weak model
$N$ \ch{of $\PS$} such that $N \sqsupset_{Max} M$. 
The set of maximal weak models of $\PS$  will be
denoted as $\MAXWM(\PS)$.~\hfill $\Box$
\end{definition}

\ch{
	In the Max Weak Model Semantics peers import maximal sets of facts
	not violating integrity constraints. Therefore,  each peer of the system can be thought as an \textit{integration resource}. 
	We will show that a locally consistent P2P system always admits a maximal weak model while a locally inconsistent P2P system not always has this property. A generalization
	of our semantics that guarantees the existence of at least a model even for locally inconsistent P2P system 
	will be presented in a following section.
}

\begin{example}\label{Ex3-maximal_cont}
In Example \ref{Ex3-maximal} the maximal weak models are
$M_2$ and $M_3$.
~\hfill $\Box$
\end{example}

\vspace*{3mm}
The Max Weak Model Semantics easily allows to express a classical problem, the \textit{three- 
colorability problem}, as follows.

\begin{example}\label{Three-colorability-example}
\chn{
	\emph{Three-colorability.} We are given two peers: $\PP_1$,
	containing a set of nodes and a set of colors which are defined by the unary relations $1\:node$ and $1\:color$ respectively, and 
	$\PP_2$, containing a set of edges defined by the binary relation $2\:edge$, the mapping rule:
}
\[
\begin{array}{ll}
2\:colored(X,C)    & \leftharpoonup 1\:node(X), 1\:color(C)
\end{array}
\]
and the integrity constraints:
\[
\begin{array}{ll}
\leftarrow 2\:colored(X,C_1),\ 2\:colored(X,C_2),\ C_1 \neq C_2 \\
\leftarrow 2\:edge(X,Y),\ 2\:colored(X,C),\ 2\:colored(Y,C) \\
\end{array}
\]
stating, respectively, that a node cannot be colored with two different colors and
two connected nodes cannot be colored with the same color.
The mapping rule states that the node $x$ can be colored with the color $c$,
only if in doing this no  constraint is violated, that is if the node $x$ is colored with a unique color
and there is no adjacent node colored with the
same color.
Each maximal weak model computes a maximal subgraph
which is three-colorable.~\hfill $\Box$
\end{example}

The following proposition shows an important property of
relation $\sqsupseteq_{Max}$.
\begin{proposition}
\chn{For any maximal P2P system $\PS = \D \cup \LP \cup \MP \cup \IC$ s.t.
	no negation occurs in $\LP $}, $\sqsupseteq_{Max}$ defines a partial order on
the set of weak models of $\PS$. 
\end{proposition}
\textbf{Proof.}
We prove that relation $\sqsupseteq_{Max}$ is antisymmetric and transitive.
\begin{itemize}
\item \emph{(Antisymmetry)}
Let us consider two weak models $M$ and $N$ in $\WM(\PS)$. We prove that if  $M\sqsupseteq_{Max}N$ and
$N\sqsupseteq_{Max}M$, then $M=N$.\ 
As \ $M\sqsupseteq_{Max}N$, \ then \ $M[\MP]\supseteq N[\MP]$. Similarly, \ as $N[\MP]\sqsupseteq_{Max}M[\MP]$, then $N[\MP]\supseteq M[\MP]$. It follows that $N[\MP]=M[\MP]$. 
\chn{
As $M$ and $N$ are weak models, by Proposition \ref{weak-model-positive-lp},
$\{M\} = \MM(St(\PS_M))$ and $\{N\} = \MM(St(\PS_N))$. Moreover, as $N[\MP]=M[\MP]$, 
$St(\PS_M)=St(\PS_N)$. 
It follows that $M=N$.}\\

\item \emph{(Transitivity)}
Let us consider three weak models $M$, $N$ and $S$ in $\WM(\PS)$. 
We prove that if  $M\sqsupseteq_{Max}N$ and
$N\sqsupseteq_{Max}S$, then $M\sqsupseteq_{Max}S$.

As $M\sqsupseteq_{Max}N$, then $M[\MP]\supseteq N[\MP]$. Similarly, as $N[\MP]\sqsupseteq_{Max}S[\MP]$, then $N[\MP]\supseteq S[\MP]$. It follows that $M[\MP]\supseteq S[\MP]$ and then
$M\sqsupseteq_{Max}S$.
\end{itemize}

~\hfill $\Box$

\chrr{
If the standard rules of a P2P system contain negation, in general $\sqsupseteq_{Max}$ is not antisymmetric. To prove it, let's consider a P2P system only containing a peer $P_1$, with the standard rules
$1\:p\leftarrow not\ 1\:q$ and 
$1\:q\leftarrow not\ 1\:p$.
The P2P system admits two weak models: $M=\{1\:p\}$ and
$N=\{1\:q\}$. As $MP[M]=MP[N]=\emptyset$, it follows that $M\sqsupseteq_{Max} N$ and $N\sqsupseteq_{Max} M$, but
$M\neq N$.
}

The next theorem shows that consistent maximal P2P systems always admit maximal weak models.
\begin{theorem} \label{th-pwm}
For every locally consistent maximal P2P system,  $\MAXWM(\PS)~\neq~\emptyset$.
\end{theorem}
\textbf{Proof.}
\ch{
Let us consider a set $M$ such that 
$\{M\}\in \MM( \D \cup \LP \cup \IC)$, that is a minimal model of a P2P system obtained from $\PS$ by deleting all mapping rules.
As $\PS$ is locally consistent, such a model exists.
Let $\Pi$ be the logic program obtained by deleting from $ground(\D\cup\LP\cup\IC)$ all peer rules whose body is false w.r.t. $M$ and by removing from the remaining rules the negative literals (observe that, they are \textit{true} w.r.t. $M$). $\Pi$ is an Horn program and admits only one minimal model. This minimal model has to be $M$.
Moreover, as $M$ does not contain any mapping atoms,
 $\Pi=St(\PS^M)$.
It follows that $\{M\}=\MM(St(\PS^M))$. This means that
$M$ is a weak model for $\PS$. As there is at least a weak model of $\PS$, then $\MAXWM(\PS)$ $~\neq~\emptyset$.
~\hfill $\Box$
}

\ch{
If a P2P system contains at least a locally inconsistent peer, the Max Weak Model semantics does not
guarantee the existence of a maximal weak model.
\begin{example}\label{inconsistentP2Psystem}
	Let us consider a P2P system containing only peer $\PP_1=\<\{1\:a,1\:b\},\emptyset,\emptyset,\{\leftarrow 1\:a,1\:b\}\>$. Clearly, $\PP_1$ is locally inconsistent and there is no way to import mapping atoms able to restore its consistency.
	Observe that, the only way to make the peer consistent is to remove at least one fact from its local database.
	In the following, we present an extension of our framework allowing deletions of facts from local databases.
	~\hfill $\Box$
\end{example}
}

\ch{\subsubsection{An Alternative Characterization of the Max Weak Models Semantics}\label{RewMax}}
In this section, we present an alternative characterization of the Max Weak Model Semantics based on the rewriting of
mapping rules into prioritized rules \cite{Bre-etal*03,SakIno00}.\\

\begin{definition}
	\textsc{[Rewriting of a Maximal P2P System into a Prioritized Logic Program].}
Given a maximal P2P system $\PS = \D \cup \LP \cup \MP \cup \IC$ and a maximal mapping rule $r \!=i\:p(x) \leftharpoonup \B$, then:
\begin{itemize}
	\item
	$Rew(r)$ denotes the pair
	$(i\:p(x) \oplus\, i\:p'(x) \leftarrow \B,$ \
	$i\:p(x) \succeq i\:p'(x))$,
	consisting of a disjunctive mapping rule and a priority statement,
	\item
	$Rew(\MP) = ( \{ Rew(r)[1] |$ $r \in \MP\}, \{ Rew(r)[2] |$ $\,r \in \MP\})$
	and
	\item
	$Rew(\PS) = ( \D \cup \LP \cup Rew(\MP)[1] \cup \IC,$ $ Rew(\MP)[2])$.\hfill$\Box$
\end{itemize}
\end{definition}
In the above definition the atom $i\:p(x)$ (resp. $i\:p'(x)$) means that
the fact $i\:p(x)$ is imported (resp. not imported) in the~peer $\PP_i$.

Intuitively, the rewriting  of the maximal mapping rule states that 
if $\B$  is \emph{true} in the \emph{source peer} then
two alternative actions can be performed in the \emph{target peer}: $i\:p(x)$ can be either imported or not imported; but the presence of the priority statement $i\:p(x) \succeq i\:p'(x))$ establishes that the action of \emph{importing $i\:p(x)$} is preferable over the
action of \emph{not importing $i\:p(x)$}.

\begin{example}
\label{Ex3-maximal_cont}
The rewriting of the P2P system  in Example \ref{Ex3-maximal} is: \\ 
$
Rew(\PS)=  (\{2\:q(a),\ 2\:q(b),$\\
\hspace*{19mm}$\ 1\:p(X)\ \oplus\  1\:p'(X)\leftarrow 2\:q(X),$\\
\hspace*{19mm}$\leftarrow 1\:p(X),\ 1\:p(Y),\ X\neq Y\},$ \\
\hspace*{19mm}$\{1\:p(X)\succeq 1\:p'(X\}).
$
\\
\noindent $Rew(\PS)[1]$ has three stable models: \\
$M_0 = \{ 2\:q(a),\ 2\:q(b),\ 1\:p'(a),\ 1\:p'(b) \}$, \\
$M_1 = \{ 2\:q(a),\ 2\:q(b),\ 1\:p(a),\  1\:p'(b) \}$,  \\
$M_2 = \{ 2\:q(a),\ 2\:q(b),\ 1\:p'(a),\ 1\:p(b) \}$.\\ \\
The set of preferred stable models are $\{M_1,M_2\}$.
\hfill $\Box$
\end{example}

\begin{example}
The rewriting of the mapping rule of Example
\ref{Three-colorability-example} consists of the 
rule:
$$
2\:colored(X,C)\   \oplus\ 2\:colored'(X,C)    \leftarrow 1\:node(X),1\:color(C) 
$$
and the preference:
$$
2\:colored(X,C)    \succeq 2\:colored'(X,C)     
$$
\hfill $\Box$
\end{example}
\noindent
Given a maximal P2P system $\PS$ and a preferred stable model $M$ for $Rew(\PS)$,
we denote with $St(M)$ the subset of non-primed atoms of $M$ and
we say that $St(M)$ is a preferred stable model of $\PS$.
\ch{
We denote the set of preferred stable models of $Rew(\PS)$ as $\PSM(\PS)$.}
The following theorem shows the equivalence of
preferred stable models and maximal weak models.
\\ \\
\normalsize
\begin{theorem}
		\textsc{[Equivalence Between Preferred Stable Models and Maximal Weak Models].}
For every maximal P2P system $\PS$,
$PSM(\PS)=\MAXWM(\PS)$.

\end{theorem}
\textbf{Proof.} 
\begin{enumerate}
\item ($PSM(\PS)\subseteq \MAXWM(\PS)$)\\
Let $M\in PSM(\PS)$ and $N=St(M)$. First we prove that $N$ is a weak model. Let us consider a ground mapping rule $m$ and its rewriting $Rew(m)$.
The rule $Rew(m)[1]=A \oplus A'\leftarrow \B$  is equivalent to the rules $r=A\leftarrow \B \wedge not\ A'$, $r'=A'\leftarrow \B\wedge not\ A$ and the constraint $\leftarrow A,A'$. 
There are three cases:
\begin{itemize}
	\item 
	$A,A'\not\in M$. In this case, $M\not\models \B$. Then the bodies of $r$ and $r'$ are false and so $r,r'\not\in (Rew(\PS)[1])^M$.
	\item 
	$A\in M$ and $A'\not\in M$. In this case the body of $r'$ is false and $r'\not\in (Rew(\PS)[1])^M$. Moreover, $A\leftarrow \B \in  (Rew(\PS)[1])^M$. 
	\item 
	$A'\in M$ and $A\not\in M$. In this case
	the body of $r$ is false and $r\not\in  (Rew(\PS)[1])^M$. Moreover, $A'\leftarrow \B \in  (Rew(\PS)[1])^M$.
\end{itemize}


Then, by construction, we have that $ (Rew(\PS)[1])^M=St(\PS^{N})\cup\{A'\leftarrow \B\ |\ A'\in M\ \wedge\ A\leftharpoonup \B\in ground(\PS)\}$. 

\noindent
We have that:
\begin{itemize}
	\item 
	The \ minimal \  model \ of \  $(Rew(\PS)[1])^M$ \ is \ $M$, \ as \  $M$ \ is \  a \ stable \ model
	of \ \ $(Rew(\PS)[1])^M$;
	\item $M=N\cup \{A'\ |\ A'\in M\}$;
	\item
	Non primed atoms $A$ can be only inferred by rules in $St(\PS^{N})$ and 
	\item
	No primed atom $A'$ occurs in the body of any rule of $St(\PS^{N})$. 
\end{itemize} 
Therefore, the minimal model of $St(\PS^{N})$ is $N$ and $N$ is a weak model of $\PS$.\\ 
Now we prove that $N$ is a maximal weak model of $\PS$. Let us assume by contradiction that there is a weak model $L$ such that $L[\MP]\supset N[\MP]$. Then the ground mapping rules that will be deleted from $ground(\PS)$ to derive $\PS^L$ are a subset of those that will be deleted to to derive $\PS^N$.


Let us consider the set $K=L\cup \{A'\ |\ A\not\in L\ \wedge\ A \leftharpoonup \B \in ground(\PS)\ \wedge\ L\models \B\}$. By construction, $K$ is the minimal model of $Rew(\PS)[1]^{K}$. Then $K$ is a stable model of $Rew(\PS)[1]$. 
Observe that, must exist two atoms $A\in K$ and $A'\in M$ and, by construction, there cannot exist two atoms $B\in M$ and $B'\in K$. Moreover, $ground(Rew(\PS)[2])$ contains the preference $A\succ A'$.  Therefore, $K \sqsupset M$ and $M$ is not a preferred stable model of $Rew(\PS)$. This is a contradiction. \\

\item ($PSM(\PS)\supseteq \MAXWM(\PS)$) \\
Let $N\in \MAXWM(\PS)$ and $M=N\cup \{A'\ |\ A\not\in N\ \wedge\ A \leftharpoonup \B \in ground(\PS)\ \wedge\ N\models \B\}$. First we prove that $M$ is a stable model of $Rew(\PS)[1]$ i.e. it is the minimal model
of $(Rew(\PS)[1])^M$.

By construction, 

$(Rew(\PS)[1])^M= St(\PS^{N}) \ \cup \ \{A'\leftarrow \B\ |\ A\not\in N\ \wedge\ A\leftharpoonup \B\in ground(\PS)\ \wedge\ N\models \B\}$.

\noindent
We have that:
\begin{itemize}
	\item 
	The minimal model of $St(\PS)^{N}$ is $N$, as $N$ is a weak model of $\PS$;
	\item 
	The minimal model of $\{A'\leftarrow \B\ |\ A\not\in N\ \wedge\ A\leftharpoonup \B\in ground(\PS)\ \wedge\ N\models \B\}$ is 
	$M\setminus N$;
	\item
	Non primed atoms $A$ can be only inferred by rules in $St(\PS^{N})$ and 
	\item
	No primed atom $A'$ occurs in the body of any rule of $St(\PS^{N})$. 
\end{itemize} 
Therefore, the \ minimal \ model of $(Rew(\PS)[1])^M$ is \ $M$ and $M$ \ is a stable \ model of $Rew(\PS)[1]$.\\ 

Now we prove that $M$ is a preferred stable model for $Rew(\PS)$. Let us assume by contradiction that there is a stable model $L$ for $Rew(\PS)$ s.t. $L\sqsupset M$. From point 1. and preferences in $Rew(\PS)[2]$, we have that 
$St(L)$ is a weak model for $\PS$ and $St(L)[\MP]\supset St(M)[\MP]$, that is $St(M)$ is not a maximal weak model for $\PS$.
This is a contradiction.	
\end{enumerate}
~\hfill $\Box$

\chn{
This characterization of the Max Weak Model Semantics
makes evident that \textit{importing} a mapping atom is preferable over \textit{not importing} it and provides a computational mechanism allowing to
derive the maximal weak models of a maximal P2P system.
}
\begin{example}
Consider the P2P system of Example \ref{Ex3-maximal_cont}, we have: \\ $\PSM(\PS)=\{\{2\:q(a),\ 2\:q(b),\ 1\:p(a)\},
	\{2\:q(a),\ 2\:q(b),\ 1\:p(b)\}\}$.
	~\hfill $\Box$
\end{example}
\chn{
This example shows that the preferred stable models of $PS$ coincide with its maximal weak models.
}

\subsection{Min Weak Model Semantics}
\label{minimal}
\ch{
In \cite{CarZum12} the authors introduced the \emph{Min Weak Model Semantics}: a peer can be locally inconsistent and the P2P system it joins  provides support to restore its consistency.
The basic idea, yet very simple, is the following: 
 an inconsistent peer, in the interaction with
other peers, just imports the missing part of its local database which is \textit{correct}, but \textit{incomplete}.\\
\noindent
The proposal of the Min Weak Model Semantics stems from the observations
that in real world P2P systems, peers often use the available import mechanisms
to extract knowledge from the rest of the system only if this knowledge
is strictly needed to repair inconsistencies of the system.}

A \emph{minimal mapping rule} (see Definition \ref{PeerDef}) is of the form $H  \leftharpoondown
\B$. Intuitively, $H \leftharpoondown \B$ means that if the body
conjunction $\B$ is \emph{true} in the source peer the atom $H$ is imported in the target peer (that is $H$ is \emph{true} in the
target peer) only if it implies (directly or indirectly) the
satisfaction of some constraints that otherwise would be violated.
\ch{
	In this section, we assume that all mapping rules of a P2P system $\PS=\<D,LP,MP,IC\>$ are minimal mapping rules i.e. $\PS$ is a minimal P2P system.
}

\begin{definition}
\textsc{[Minimal Weak Model].} 
\ch{
Given a minimal P2P system $\PS$ and two weak models $M$ and $N$ of $\PS$,} 
$M$ is said \emph{min-preferable}
to $N$, and is denoted as $M \sqsupseteq_{Min} N$, if $M[\MP] \subseteq N[\MP]$.
Moreover, if $M \sqsupseteq_{Min} N$ and $N \not\sqsupseteq_{Min} M$ then $M
\sqsupset_{Min} N$.
A weak model $M$ \ch{of $\PS$} is said to be  \emph{minimal} if there is no weak model
$N$ \ch{of $\PS$} such that $N \sqsubset_{Min} M$.
The set of minimal weak models  will be
denoted by $\MINWM(\PS)$.~\hfill $\Box$
\end{definition}

Next example will clarify the concept of minimal weak model.
\ch{
	\begin{example}\label{Motivating-Example-minimal-Cont2}
		Consider the P2P system $\PS$ presented in Example \ref{Motivating-Example-minimal}.
		The weak models of the system are:\\ \\
		$M_1=\{1\:vendor(dan,laptop),\ 1\:vendor(bob,laptop),\ 2\:order(laptop),$\\ 
		\hspace*{11mm}$2\:supplier(dan,laptop),\ 2\:available(laptop)\}$,\\ \\
		$M_2=\{1\:vendor(dan,laptop),\ 1\:vendor(bob,laptop),\ 2\:order(laptop),$\\ 
		\hspace*{11mm}$2\:supplier(bob,laptop),\ 2\:available(laptop)\}$ and\\ \\
		$M_3=\{1\:vendor(dan,laptop),\ 1\:vendor(bob,laptop),\ 2\:order(laptop),$\\ 
		\hspace*{11mm}$2\:supplier(dan,laptop),\ 2\:supplier(bob,laptop),\ 2\:available(laptop)\}$,\\ \\
		\noindent
		whereas the minimal weak models are
		$M_1$ and $M_2$ because they contain minimal subsets of mapping atoms
		(resp. $2\:\{supplier(dan,laptop)\}$ and $\{2\:supplier(bob,$ $laptop)\}$).
		~\hfill~$\Box$
	\end{example}
}

We observe that, adopting the Min Weak Model Semantics, if each peer of a P2P system is locally consistent then
no mapping atom is inferred.
Clearly, not always a minimal weak model exists. This happens when
there is at least a peer which is locally inconsistent and there is no way
to import mapping atoms that could repair its local database so that
its consistency can be restored.
\ch{
	\begin{example}
		Let us consider the simple P2P system presented in Example \ref{inconsistentP2Psystem}. Also adopting the Min Weak Model Semantics, $\PS$ does not admit any minimal weak model. 	~\hfill~$\Box$
	\end{example}
}
\ch{
	It is important to observe that a peer uses its minimal mapping rules to import minimal sets of atoms allowing the satisfaction of integrity constraints belonging not only to it but also to other peers.
	\begin{example}
		\label{ExampleGlobalInconsistency}
		\begin{figure}[h]
			\centering
			\includegraphics[width=0.9\textwidth]{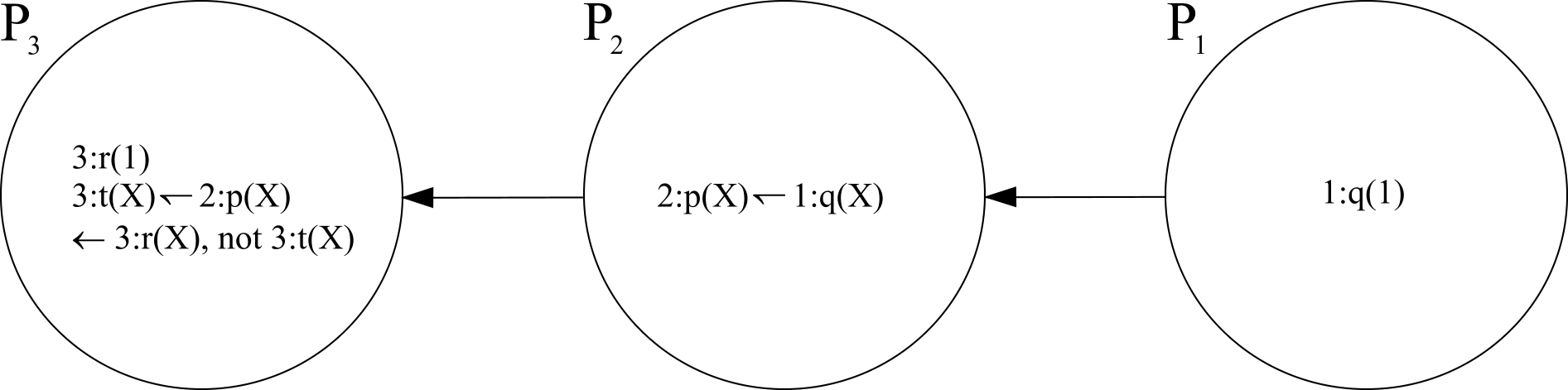}
			\caption{A P2P System with maximal and minimal mapping rules}
			\label{fig-GlobalInconsistency}
		\end{figure}
		Consider the P2P system depicted in Figure \ref{fig-GlobalInconsistency}. 
		\begin{itemize}
			\item
			Peer
			$\PP_1$ contains the 
			fact $1\:q(1)$
			\item
			Peer $\PP_2$ contains the minimal mapping rule\ $2\:p(X)\leftharpoondown 1\:q(X)$
			\item
			Peer $\PP_3$ contains the fact $3\:r(1)$,
				 the minimal mapping rule $1\:t(X)\leftharpoondown 2\:p(X)$ and
				the integrity constraint 
				$\leftarrow 3\:r(X),\ not \ 3\:t(X)$.			
		\end{itemize}
		Peer $\PP_2$ imports the atom $2\:p(1)$ from $\PP_1$ to guarantee the satisfaction of an integrity constraint belonging to $\PP_3$. Min Weak Model Semantics assigns to the system its unique minimal weak model: $\{1\:q(1),\ 2\:p(1),3\:r(1),3\:t(1)\}$. 	~\hfill~$\Box$
	\end{example} 
}
The following proposition shows an important property of relation $\sqsupseteq_{Min}$.
\begin{proposition}
\chn{For any maximal P2P system $\PS = \D \cup \LP \cup \MP \cup \IC$ s.t.
	no negation occurs in $\LP $}, $\sqsupseteq_{Min}$ defines a partial order on
the set of weak models of $\PS$. 
\end{proposition}
\textbf{Proof.}
We prove that relation $\sqsupseteq_{Min}$ is antisymmetric and transitive.
\begin{itemize}
\item \emph{(Antisymmetry)}
Let us consider the weak models $M$ and $N$ in $\PS$. We prove that if  $M\sqsupseteq_{Min}N$ and
$N\sqsupseteq_{Min}M$, then $M=N$.
As $M\sqsupseteq_{Min}N$, then $M[\MP]\subseteq N[\MP]$. Similarly, as $N[\MP]\sqsupseteq_{Min}M[\MP]$, then $N[\MP]\subseteq M[\MP]$. It follows that $N[\MP]=M[\MP]$. 
\chn{
As $M$ and $N$ are weak models, by Proposition \ref{weak-model-positive-lp},
$\{M\} = \MM(St(\PS_M))$ and $\{N\} = \MM(St(\PS_N))$. Moreover, as $N[\MP]=M[\MP]$, 
$St(\PS_M)=St(\PS_N)$.
It follows that $M=N$. }\\

\item \emph{(Transitivity)}
Let us consider the weak models $M$, $N$ and $S$ in $\PS$. 
We prove that if  $M\sqsupseteq_{Min}N$ and
$N\sqsupseteq_{Min}S$, then $M\sqsupseteq_{Min}S$.

As $M\sqsupseteq_{Min}N$, then $M[\MP]\subseteq N[\MP]$. Similarly, as $N[\MP]\sqsupseteq_{Min}S[\MP]$, then $N[\MP]\subseteq S[\MP]$. It follows that $M[\MP]\subseteq S[\MP]$ and then
$M\sqsupseteq_{Min}S$.
\end{itemize}
~\hfill $\Box$

In the Min Weak Model Semantics a  peer may import an atom from a neighbor peer even if such atom is not needed to repair its own inconsistency, but is needed to restore the consistency of a different peer. In this way, the system behaves in a global way. 

\ch{\subsubsection{An Alternative Characterization of the Min Weak Models Semantics}}\label{RewMin}
Similarly to Section \ref{RewMax}, here we present an alternative characterization of the Min Weak Model Semantics
based on the rewriting of
mapping rules into  prioritized rules \cite{Bre-etal*03,SakIno00}. \\

\begin{definition}
	\textsc{[Rewriting of a Maximal P2P System into a Prioritized Logic Program].}
Given a minimal P2P system $\PS$ and a mapping rule $r \!=i\:p(x) \leftharpoondown \B$, then:
\begin{itemize}
	\item
	$Rew(r)$ denotes the pair
	$(i\:p(x) \oplus\, i\:p'(x) \leftarrow \B,$ \
	$i\:p'(x) \succeq i\:p(x))$,
	consisting of a disjunctive mapping rule and a priority statement,
	\item
	$Rew(\MP) = ( \{ Rew(r)[1] |$ $r \in \MP\}, \{ Rew(r)[2] |$ $\,r \in \MP\})$
	and
	\item
	$Rew(\PS) = ( \D \cup \LP \cup Rew(\MP)[1] \cup \IC,$ $ Rew(\MP)[2])$.\hfill$\Box$
\end{itemize}
\end{definition}
In the above definition, the atom $i\:p(x)$ (resp. $i\:p'(x)$) means that
the fact $i\:p(x)$ is imported (resp. not imported) in the peer $\PP_i$.

\noindent

Intuitively, the rewriting  of the mapping rule states that 
if $\B$  is \emph{true} in the \emph{source peer} then
two alternative actions can be performed in the \emph{target peer}: $i\:p(x)$ can be either imported or not imported; but the presence of the priority statement $i\:p'(x) \succeq i\:p(x))$ establishes that the action of \emph{not importing $i\:p(x)$} is preferable over the
action of \emph{importing $i\:p(x)$}.

\ch{
\begin{example}
Consider again the system presented in Example \ref{Motivating-Example-minimal}.
The rewriting of the system is: \\ \\
\noindent
$Rew(\PS)=\{1\:vendor(dan,laptop),1\:vendor(bob,laptop),2\:order(laptop),$\\
\hspace*{19mm}$2\:supplier(X,Y)\oplus 2\:supplier'(X,Y)\leftarrow 1\:vendor(X,Y),$\\
\hspace*{19mm}$1\:available(Y)\leftarrow 1\:supplier(X,Y),$\\
\hspace*{19mm}$\leftarrow 1\:order(X), \  not\ 1\:available(X)\},$\\ \
\hspace*{19mm}$\{1\:supplier'(X,Y)\succeq 1\: supplier(X,Y)\}).$\\ \\
\noindent $Rew(\PS)[1]$ has three stable models: \\ \\
$M_1=\{1\:vendor(dan,laptop),1\:vendor(bob,laptop),2\:order(laptop),$\\ 
\hspace*{10mm} $2\:supplier(dan,laptop),2\:supplier'(bob,laptop),2\:available(laptop)\}$,\\ \\
$M_2=\{1\:vendor(dan,laptop),1\:vendor(bob,laptop),2\:order(laptop),$\\
\hspace*{10mm} $2\:supplier'(dan,laptop),2\:supplier(bob,laptop),2\:available(laptop)\}$,\\ \\
$M_3=\{1\:vendor(dan,laptop),1\:vendor(bob,laptop),2\:order(laptop),$\\ 
\hspace*{10mm} $2\:supplier(dan,laptop),2\:supplier(bob,laptop),2\:available(laptop)\}$.\\ \\
\noindent
The preferred stable models are $M_1$ and $M_2$.
\hfill $\Box$
\end{example}
}
The following theorem shows the equivalence of
preferred stable models and minimal weak models.

\normalsize
\begin{theorem}
		\textsc{[Equivalence Between Preferred Stable Models and Minimal Weak Models].}
For every minimal P2P system $\PS$,
$PSM(\PS)=\MINWM(\PS)$.
\end{theorem}
\textbf{Proof.}
\begin{enumerate}
\item ($PSM(\PS)\subseteq \MINWM(\PS)$)

Let $M\in PSM(\PS)$ and $N=St(M)$. First we prove that $N$ is a weak model. Let us consider a ground mapping rule $m$ and its rewriting $Rew(m)$.
The rule $Rew(m)[1]=A \oplus A'\leftarrow \B$  is equivalent to the rules $r=A\leftarrow \B \wedge not\ A'$, $r'=A'\leftarrow \B\wedge not\ A$ and the constraint $\leftarrow A,A'$. 
There are three cases:
\begin{itemize}
	\item 
	$A,A'\not\in M$. In this case, $M\not\models \B$. Then the bodies of $r$ and $r'$ are false and so $r,r'\not\in (Rew(\PS)[1])^M$.
	\item 
	$A\in M$ and $A'\not\in M$. In this case the body of $r'$ is false and $r'\not\in (Rew(\PS)[1])^M$. Moreover, $A\leftarrow \B \in  (Rew(\PS)[1])^M$. 
	\item 
	$A'\in M$ and $A\not\in M$. In this case
	the body of $r$ is false and $r\not\in  (Rew(\PS)[1])^M$. Moreover, $A'\leftarrow \B \in  (Rew(\PS)[1])^M$.
\end{itemize}


Then, by construction, we have that $ (Rew(\PS)[1])^M=St(\PS^{N})\cup\{A'\leftarrow \B\ |\ A'\in M\ \wedge\ A\leftharpoonup \B\in ground(\PS)\}$. 

\noindent
We have that:
\begin{itemize}
	\item 
	The \ minimal \  model \ of \  $(Rew(\PS)[1])^M$ \ is \ $M$, \ as \  $M$ \ is \  a \ stable \ model
	of \ \ $(Rew(\PS)[1])^M$;
	\item $M=N\cup \{A'\ |\ A'\in M\}$;
	\item
	Non primed atoms $A$ can be only inferred by rules in $St(\PS^{N})$ and 
	\item
	No primed atom $A'$ occurs in the body of any rule of $St(\PS^{N})$. 
\end{itemize} 
Therefore, the minimal model of $St(\PS^{N})$ is $N$ and $N$ is a weak model of $\PS$.\\ 
Now we prove that $N$ is a minimal weak model of $\PS$. Let us assume by contradiction that there is a weak model $L$ such that $L[\MP]\subset N[\MP]$. Then the ground mapping rules that will be deleted from $ground(\PS)$ to derive $\PS^L$ are a superset of those that will be deleted to to derive $\PS^N$.


Let us consider the set $K=L\cup \{A'\ |\ A\not\in L\ \wedge\ A \leftharpoondown \B \in ground(\PS)\ \wedge\ L\models \B\}$. By construction, $K$ is the minimal model of $Rew(\PS)[1]^{K}$. Then $K$ is a stable model of $Rew(\PS)[1]$. 
Observe that must exist two atoms $A'\in K$ and $A\in M$ and, by construction, there cannot exist two atoms $B'\in M$ and $B\in K$. Moreover, $ground(Rew(\PS)[2])$ contains the preference $A'\succ A$.  Therefore, $K \sqsupset M$ and $M$ is not a preferred stable model of $Rew(\PS)$. This is a contradiction. \\

\item($PSM(\PS)\supseteq \MINWM(\PS)$)

Let $N\in \MINWM(\PS)$ and $M=N\cup \{A'\ |\ A\not\in N\ \wedge\ A \leftharpoondown \B \in ground(\PS)\ \wedge\ N\models \B\}$. First we prove that $M$ is a stable model of $Rew(\PS)[1]$ i.e. it is the minimal model
of $(Rew(\PS)[1])^M$.

By construction, $(Rew(\PS)[1])^M=St(\PS^{N})\cup\{A'\leftarrow \B\ |\ A\not\in N\ \wedge\ A\leftharpoondown \B\in ground($ $\PS)\ \wedge\ N\models \B\}$.
We have that:
\begin{itemize}
	\item 
	The minimal model of $St(\PS)^{N}$ is $N$, as $N$ is a weak model of $\PS$;
	\item 
	The minimal model of $\{A'\leftarrow \B\ |\ A\not\in N\ \wedge\ A\leftharpoondown \B\in ground($ $\PS)\ \wedge\ N\models \B\}$ is 
	$M\setminus N$;
	\item
	Non primed atoms $A$ can be only inferred by rules in $St(\PS^{N})$ and 
	\item
	No primed atom $A'$ occurs in the body of any rule of $St(\PS^{N})$. 
\end{itemize} 
Therefore, the \ minimal \ model \ of $(Rew(\PS)[1])^M$ is $M$ \ and $M$ is a stable model of $Rew(\PS)[1]$.\\ 

Now we prove that $M$ is a preferred stable model for $Rew(\PS)$. Let us assume by contradiction that there is a stable model $L$ for $Rew(\PS)$ s.t. $L\sqsupset M$. From point 1. and preferences in $Rew(\PS)[2]$, we have that 
$St(L)$ is a weak model for $\PS$ and $St(L)[\MP]\subset St(M)[\MP]$, that is $St(M)$ is not a minimal weak model for $\PS$.
This is a contradiction.~\hfill $\Box$	
\end{enumerate}

\chn{
	This characterization of the Min Weak Model Semantics
	makes evident that \textit{not importing} a mapping atom is preferable over \textit{importing} it and provides a computational mechanism allowing to
	derive the minimal weak models of a minimal P2P system.
}

\ch{\subsection{Max-Min Weak Model Semantics}}


This section presents  a unified semantics for P2P systems, the Max-Min Weak Model Semantics,  that represents a generalization of those introduced in
Section \ref{maximal} and Section \ref{minimal}. 
A peer, \ch{which can be locally inconsistent}, can use two import mechanisms for importing knowledge from other peers: maximal mapping rules to import maximal sets of mapping atoms not violating local integrity constraints and minimal mapping rules 
to restore local consistency.
These two mechanisms can be combined and used in the same peer. With this semantics, a peer
can consider each of its neighbors as a resource
used either to enrich (integrate) or to fix (repair) its knowledge,
so as to define a kind of \emph{integrate\-repair} strategy. \\

\begin{example}
\label{Motivating-example-Max-Min-Cont1}

Consider again the P2P system presented in Example \ref{Motivating-example-Max-Min}.
As we observed, peer $\PP_3$ is locally consistent.
It imports from
$\PP_1$ all the orders that can be satisfied by suppliers 
imported from peer $\PP_2$. 
Moreover, a minimum set of suppliers will be imported in $\PP_3$.

The fact $3\:order(laptop)$ will be imported in $\PP_3$ from  $\PP_1$ because there is at least a supplier of the object `laptop' that can be imported 
in $\PP_3$ from  $\PP_2$. Instead, there is no way to import a supplier of the object `monitor'. Therefore, the fact $3\:order(monitor)$ will not be imported in
$\PP_3$. Finally, there are two possible ways to import a supplier of
the object `laptop': importing from $\PP_1$ either the fact $3\:supplier(dan,
laptop)$ or the fact $3\:supplier(bob, laptop)$.
 ~\hfill~$\Box$
\end{example}


\begin{definition}
\textsc{[Max-Min Weak Model.]} 
\ch{Given a P2P system $\PS$} and two weak models $M$ and $N$ \ch{of $\PS$}, we say that $M$ is \emph{max-min-preferable}
to $N$, and we write $M \sqsupseteq N$, if
\begin{itemize}
	\item
	$M[\overline{\MP}] \supset N[\overline{\MP}]$ \ \ or
	\item
	$M[\overline{\MP}] = N[\overline{\MP}]$ and
	$M[\underline{\MP}] \subseteq N[\underline{\MP}]$
\end{itemize}
Moreover, if $M \sqsupseteq N$ and $N \not\sqsupseteq M$ we write $M
\sqsupset N$.
A weak model $M$ is said to be  \emph{max-min} if there is no weak model
$N$ such that $N \sqsubset M$.
The set of max-min weak models will be
denoted by $\PWM(\PS)$.
~\hfill $\Box$
\end{definition}


\ch{ 
%
The above definition states that a weak model is a max-min weak model if it maximizes the set of atoms imported by means of maximal mapping rules while minimizing the set of atoms imported my means of minimal mapping rules (used to maintain local consistency). The approach follows the \textit{classical} and \textit{natural} strategy of enriching as much as possible the knowledge of an information source (by means of maximal mapping rules) guaranteeing consistency (by using minimal mapping rules). 
}

\ch{
	\begin{example}\label{Motivating-example-Max-Min-Cont2}
		Consider the P2P system $\PS$ presented in Example \ref{Motivating-example-Max-Min}.
		The weak models of the system are:\\ \\
		\noindent
		$M_1=\{1\:vendor(dan,laptop),1\:vendor(bob,laptop),$\\
		\hspace*{10mm}$2\:shopping(laptop),2\:shopping(monitor)\}$,\\ \\
		\noindent
		$M_2=\{1\:vendor(dan,laptop),1\:vendor(bob,laptop), 2\:shopping(laptop),$\\ 
		\hspace*{10mm}$2\:shopping(monitor),3\:supplier(dan,laptop),3\:available(laptop)\}$,\\ \\
		\noindent
		$M_3=\{1\:vendor(dan,laptop),1\:vendor(bob,laptop), 2\:shopping(laptop),$\\ 
		\hspace*{10mm}$2\:shopping(monitor), 3\:supplier(bob,laptop),3\:available(laptop)\}$,\\ \\
		\noindent
		$M_4=\{1\:vendor(dan,laptop),1\:vendor(bob,laptop),2\:shopping(laptop),$\\ 
		\hspace*{10mm}$2\:shopping(monitor),3\:supplier(dan,laptop),3\:supplier(bob,laptop),$\\ 
		\hspace*{10mm}$3\:available(laptop)\}$,\\ \\
		\noindent
		$M_5=\{1\:vendor(dan,laptop),1\:vendor(bob,laptop),2\:shopping(laptop),$\\ 
		\hspace*{10mm}$2\:shopping(monitor),3\:supplier(dan,laptop),3\:order(laptop), $\\ 
		\hspace*{10mm}$3\:available(laptop)\}$,\\ \\
		\noindent
		$M_6=\{1\:vendor(dan,laptop),1\:vendor(bob,laptop),2\:shopping(laptop),$\\ 
		\hspace*{10mm}$2\:shopping(monitor),3\:supplier(bob,laptop),3\:order(laptop),$\\ 
		\hspace*{10mm}$3\:available(laptop)\}$,\\ \\
		\noindent
		$M_7=\{1\:vendor(dan,laptop),1\:vendor(bob,laptop),2\:shopping(laptop),$\\ 
		\hspace*{10mm}$2\:shopping(monitor),3\:supplier(dan,laptop),3\:supplier(bob,laptop),$\\ 
		\hspace*{10mm}$3\:order(laptop),3\:available(laptop)\}$.\\ \\
		\noindent
		whereas the max-min weak models are $M_5$ and $M_6$.
		~\hfill~$\Box$
	\end{example}
}
%
%


The following proposition shows an important property of relation  $\sqsupseteq$.
It is easy to show that a locally consistent P2P system always admits a max-min weak model.

\begin{proposition}
\chn{For any maximal P2P system $\PS = \D \cup \LP \cup \MP \cup \IC$ s.t.
	no negation occurs in $\LP $}, $\sqsupseteq$ defines a partial order on
the set of weak models of $\PS$.
\end{proposition}
\textbf{Proof.}
We prove that relation $\sqsupseteq$ is antisymmetric and transitive.
\begin{itemize}
\item \emph{(Antisymmetry)}
Let us consider the weak models $M$ and $N$ in $\PS$. We prove that if  $M\sqsupseteq N$ and
$N\sqsupseteq M$, then $M=N$.
As $M\sqsupseteq N$, then $M[\overline{\MP}]\supset N[\overline{\MP}]$ or
$M[\overline{\MP}]=N[\overline{\MP}]$ and 
$M[\underline{\MP}]\subseteq N[\underline{\MP}]$.
Similarly, as $N\sqsupseteq M$, then $N[\overline{\MP}]\supset M[\overline{\MP}]$ or
$N[\overline{\MP}]=M[\overline{\MP}]$ and 
$N[\underline{\MP}]\subseteq M[\underline{\MP}]$.
As \ the \ conditions $M[\overline{\MP}]\supset N[\overline{\MP}]$ and $N[\overline{\MP}]\supset M[\overline{\MP}]$ cannot holds at the same time, it follows that $N[\overline{\MP}]=M[\overline{\MP}]$ and 
$N[\underline{\MP}]= M[\underline{\MP}]$, that is
$N[\MP]=M[\MP]$. 
\chn{
As $M$ and $N$ are weak models, by Proposition \ref{weak-model-positive-lp},
$\{M\} = \MM(St(\PS_M))$ and $\{N\} = \MM(St(\PS_N))$. Moreover, as $N[\MP]=M[\MP]$, 
$St(\PS_M)=St(\PS_N)$. 
It follows that $M=N$.}
\item \emph{(Transitivity)}
Let us consider the weak models $M$, $N$ and $S$ in $\PS$. 
We prove that if  $M\sqsupseteq N$ and
$N\sqsupseteq S$, then $M\sqsupseteq S$.

If $M[\overline{\MP}]\supset N[\overline{\MP}]$ and
$N[\overline{\MP}]\supset S[\overline{\MP}]$, then
$M[\overline{\MP}]\supset S[\overline{\MP}]$ and so
$M\sqsupseteq S$.

If $M[\overline{\MP}]\supset N[\overline{\MP}]$,
$N[\overline{\MP}]=S[\overline{\MP}]$ and 
$N[\underline{\MP}]\subseteq S[\underline{\MP}]$,
then
$M[\overline{\MP}]$ $\supset S[\overline{\MP}]$ and so
$M\sqsupseteq S$.

If 
$M[\overline{\MP}]=N[\overline{\MP}]$, \ \
$M[\underline{\MP}]\subseteq N[\underline{\MP}]$,
and \ \
$N[\overline{\MP}]\supset S[\overline{\MP}]$, \ then  \ 
$M[$ $\overline{\MP}]$ $\supset S[\overline{\MP}]$ and so
$M\sqsupseteq S$.

If 
$M[\overline{\MP}]=N[\overline{\MP}]$, 
$M[\underline{\MP}]\subseteq N[\underline{\MP}]$,
$N[\overline{\MP}]=S[\overline{\MP}]$ and 
$N[\underline{\MP}]\subseteq S[\underline{\MP}]$,
then
$M[\overline{\MP}]=S[\overline{\MP}]$, 
$M[\underline{\MP}]\subseteq S[\underline{\MP}]$ and so
$M\sqsupseteq S$.
\end{itemize}

~\hfill $\Box$

\ch{\subsubsection{An Alternative Characterization of the Max-Min Weak Models}}\label{RewMaxMin}


Similarly to Section \ref{RewMax} and Section \ref{RewMin}, this section presents an alternative characterization of the Max-Min
Weak Model Semantics based on the rewriting of mapping rules into
prioritized rules \cite{Bre-etal*03,SakIno00}. 
Given an atom
$A=i\:p(x_1,\ldots,x_n)$ we denote as $A'$ the atom
$i\:p'(x_1,\ldots,$ $x_n)$. \\

\begin{definition}
		\textsc{[Rewriting of a P2P System into a Prioritized Logic Program].}
Given a P2P system $\PS = \D \cup \LP \cup \MP \cup \IC$
and the mapping rules $r_a \!=i_a\:\ p_a(x_a) \leftharpoonup \B_a$ and
$r_b \!=i_b\:\ p_b(x_b)\leftharpoondown \B_b$, then:
\begin{itemize}
	\item
	$Rew(r_a)$ denotes the pair \\
	$(i_a\:\ p_a(x_a) \oplus\, i_a\:\ p_a'(x_a) \leftarrow \B_a,$ \
	$i_a\:\ p_a(x_a) \succeq i_a\:\ p_a'(x_a))$,
	\vspace{3mm}
	\item
	$Rew(r_b)$ denotes the pair \\
	$(i_b\:\ p_b(x_b) \oplus\, i_b\:\ p_b'(x_b) \leftarrow \B_b,$ \
	$i_b\:\ p_b'(x_b) \succeq i_b\:\ p_b(x_b))$,
	\vspace{3mm}
	\item
	$Rew(\overline{\MP})$ denotes the pair\\ $( \{ Rew(r)[1] |$ $r \in
	\overline{\MP}\}, \{ Rew(r)[2] |$ $\,r \in \overline{\MP}\})$
	\vspace{3mm}
	\item
	$Rew(\underline{\MP})$ denotes the pair\\ $( \{ Rew(r)[1] |$ $r \in
	\underline{\MP}\}, \{ Rew(r)[2] |$ $\,r \in \underline{\MP}\})$ and
	\vspace{3mm}
	\item
	$Rew(\PS)$ denotes the prioritized logic program \\$( \D \cup \LP \cup Rew(\overline{\MP})[1]\cup Rew(\underline{\MP})[1] \cup \IC,$\\ $
	Rew(\overline{\MP})[2],$ $Rew(\underline{\MP})[2])$.\hfill $\Box$
\end{itemize}
\end{definition}
\normalsize 
\ch{
In the above definition, the atom $i_a\:\ p_a(x_a)$ (resp. $i_b\:\ p_b(x_b)$) means that the
fact $1_a\:p_a(x_a)$ is imported in peer $\PP_{i_a}$ (resp. $1_b\:p_b(x_b)$ is imported in peer $\PP_{i_b}$).
\\
Intuitively, the rewriting  of the maximal (resp. minimal) mapping rule states that 
if $\B_a$ (resp. $\B_b$) is \emph{true} in the \emph{source peer} then
two alternative actions can be performed in the \emph{target peer}: $i_a\:p_a(x_a)$ (resp. $i_b\:p_b(x_b)$) can be either imported or not imported; but the presence of the priority statement $i_a\:p_a(x_a) \succeq i_a\:p_a'(x_a)$ (resp. $i_b\:p_b'(x_b) \succeq i_b\:p_b(x_b)$) establishes that the action of \emph{importing $i_a\:p_a(x_a)$} is preferable over the
action of \emph{not importing $i_a\:p_a(x_a)$} (resp. the action of \emph{not importing $i_b\:p_b(x_b)$} is preferable over the
action of \emph{importing $i_b\:p_b(x_b)$}).
}

Observe that, $Rew(\PS)$ is a prioritized logic program with two
levels of priorities. The one applied as first models the preference
to import as much maximal mapping atoms as possible. The other one,
applied over the models selected in the first step, expresses the
preference to import as less minimal mapping atoms as possible.

\ch{
\begin{example}
Consider again the system reported in Example \ref{fig-Motivating-Example-Max-Min}. 
The rewriting of the system is: 
\\
\noindent
$Rew(\PS)= $ \\
\hspace*{12mm}$(\{1\:vendor(dan,laptop),1\:vendor(bob,laptop),$\\
\hspace*{15mm}$2\:shopping(laptop),2\:shopping(monitor)$\\
\hspace*{15mm}$3\:available(Y)\leftarrow 3\:supplier(X,Y),$\\
\hspace*{15mm}$3\:supplier(X,Y)\oplus 3\:supplier'(X,Y)\leftarrow 1\:vendor(X,Y),$\\
\hspace*{15mm}$3\:order(X)\oplus 3\:order'(X)\leftarrow 2\:shopping(X),$\\
\hspace*{15mm}$\leftarrow 3\:order(X),not\ 3\:available(X)$\}, \\ \\
\hspace*{15mm}$\{3\:order(X)\succeq 3\:order'(X)\},$ \\
\hspace*{15mm}$ \{ 3\:supplier'(X,Y)\succeq 3\:supplier(X,Y)\}).$\\
\\
\noindent The logic program has the following stable models: \\
\\
\noindent
$M_1=\{1\:vendor(dan,laptop),1\:vendor(bob,laptop),$\\
\hspace*{10mm}$2\:shopping(laptop), 2\:shopping(monitor),$\\
\hspace*{10mm}$3\:supplier'(dan,laptop),3\:supplier'(bob,laptop),$\\
\hspace*{10mm}$3\:order'(laptop),3\:order'(monitor)\}$,\\
\\
\noindent
$M_2=\{1\:vendor(dan,laptop),1\:vendor(bob,laptop),$\\
\hspace*{10mm}$2\:shopping(laptop),2\:shopping(monitor),$\\
\hspace*{10mm}$3\:supplier(dan,laptop),3\:supplier'(bob,laptop),$\\
\hspace*{10mm}$3\:order'(laptop),3\:order'(monitor),3\:available(laptop)\}$,\\
\\
\noindent
$M_3=\{1\:vendor(dan,laptop),1\:vendor(bob,laptop),$\\
\hspace*{10mm}$2\:shopping(laptop),2\:shopping(monitor),$\\
\hspace*{10mm}$3\:supplier'(dan,laptop),3\:supplier(bob,laptop),$\\
\hspace*{10mm}$3\:order'(laptop),3\:order'(monitor),3\:available(laptop)\}$,\\
\\
\noindent
$M_4=\{1\:vendor(dan,laptop),1\:vendor(bob,laptop),$\\
\hspace*{10mm}$2\:shopping(laptop),2\:shopping(monitor),$\\
\hspace*{10mm}$3\:supplier(dan,laptop),3\:supplier(bob,laptop),$\\
\hspace*{10mm}$3\:order'(laptop),3\:order'(monitor),3\:available(laptop)\}$,\\
\\
\noindent
$M_5=\{1\:vendor(dan,laptop),1\:vendor(bob,laptop),$\\
\hspace*{10mm}$2\:shopping(laptop),2\:shopping(monitor),$\\
\hspace*{10mm}$3\:supplier(dan,laptop),3\:supplier'(bob,laptop),$\\
\hspace*{10mm}$3\:order(laptop),3\:order'(monitor),3\:available(laptop)\}$,\\
\\
\noindent
$M_6=\{1\:vendor(dan,laptop),1\:vendor(bob,laptop),$\\
\hspace*{10mm}$2\:shopping(laptop),2\:shopping(monitor),$\\
\hspace*{10mm}$3\:supplier'(dan,laptop),3\:supplier(bob,laptop),$\\
\hspace*{10mm}$3\:order(laptop),3\:order'(monitor),3\:available(laptop)\}$,\\
\\
\noindent
$M_7=\{1\:vendor(dan,laptop),1\:vendor(bob,laptop),$\\
\hspace*{10mm}$2\:shopping(laptop),2\:shopping(monitor),$\\
\hspace*{10mm}$3\:supplier(dan,laptop),3\:supplier(bob,laptop),$\\
\hspace*{10mm}$3\:order(laptop),3\:order'(monitor),3\:available(laptop)\}$,\\
\\
\noindent
The preferred stable models are $M_5$ and $M_6$.
\hfill $\Box$
\end{example}
}
\noindent

Given a P2P system $\PS$ and a preferred stable model $M$ for $Rew(\PS)$
we denote with $St(M)$ the subset of non-primed atoms of $M$ and
we say that $St(M)$ is a preferred stable model of $\PS$.
We denote the set of preferred stable models of $\PS$ as $\PSM(\PS)$.

The following theorem shows the equivalence of
preferred stable models and max-min weak models.

\normalsize
\begin{theorem}
\label{equivalence}
		\textsc{[Equivalence Between Preferred Stable Models and Max-Min Weak Models].}
For every P2P system $\PS$, $PSM(\PS)= \PWM(\PS)$. 
\end{theorem}
\textbf{Proof.}
\begin{enumerate}
\item($PSM(\PS)\subseteq \PWM(\PS)$)

Let $M\in PSM(\PS)$ and $N=St(M)$. First we prove that $N$ is a weak model. Let us consider a ground mapping rule $m$ and its rewriting $Rew(m)$.
The rule $Rew(m)[1]=A \oplus A'\leftarrow \B$  is equivalent to the rules $r=A\leftarrow \B \wedge not\ A'$, $r'=A'\leftarrow \B\wedge not\ A$ and the constraint $\leftarrow A,A'$. 
There are three cases:
\begin{itemize}
	\item 
	$A,A'\not\in M$. In this case, $M\not\models \B$. Then the bodies of $r$ and $r'$ are false and so $r,r'\not\in (Rew(\PS)[1])^M$.
	\item 
	$A\in M$ and $A'\not\in M$. In this case the body of $r'$ is false and $r'\not\in (Rew(\PS)[1])^M$. Moreover, $A\leftarrow \B \in  (Rew(\PS)[1])^M$. 
	\item 
	$A'\in M$ and $A\not\in M$. In this case
	the body of $r$ is false and $r\not\in  (Rew(\PS)[1])^M$. Moreover, $A'\leftarrow \B \in  (Rew(\PS)[1])^M$.
\end{itemize}


Then, by construction, we have that $ (Rew(\PS)[1])^M=St(\PS^{N})\cup\{A'\leftarrow \B\ |\ A'\in M\ \wedge\ A\leftharpoonup \B\in ground(\PS)\}$. 

\noindent
We have that:
\begin{itemize}
	\item 
	The \ minimal \  model \ of \  $(Rew(\PS)[1])^M$ \ is \ $M$, \ as \  $M$ \ is \  a \ stable \ model
	of \ \ $(Rew(\PS)[1])^M$;
	\item $M=N\cup \{A'\ |\ A'\in M\}$;
	\item
	Non primed atoms $A$ can be only inferred by rules in $St(\PS^{N})$ and 
	\item
	No primed atom $A'$ occurs in the body of any rule of $St(\PS^{N})$. 
\end{itemize} 
Therefore, the minimal model of $St(\PS^{N})$ is $N$ and $N$ is a weak model of $\PS$.\\ 
Now we prove that $N$ is a preferred weak model of $\PS$. Let us assume by contradiction that there is a weak model $L$ such that $L[\overline{\MP}]\supset N[\overline{\MP}]$ or $L[\overline{\MP}]= N[\overline{\MP}]\ \wedge\ L[\underline{\MP}]\subset N[\underline{\MP}]$. 
Let us consider these cases:
\begin{itemize}
	\item
	($L[\overline{\MP}]\supset N[\overline{\MP}]$).
	In this case, the ground maximal mapping rules that will be deleted from $ground(\PS)$ to derive $\PS^L$ are a subset of those that will be deleted to to derive $\PS^N$.
	
	
	Let us consider the set $K=L\cup \{A'\ |\ A\not\in L\ \wedge(A\leftharpoonup \B\in ground(\PS)\ \vee\ A\leftharpoondown \B\in ground(\PS)) \ \wedge\ L\models \B\}$. By construction, it is the minimal model of $Rew(\PS)[1]^{K}$. Then $K$ is a stable model of $Rew(\PS)[1]$. 
	Observe \  that must \ exist \ two \ atoms $A\in K$ and $A'\in M$ and, by construction, there cannot exist two atoms $B\in M$ and $B'\in K$. 
	Moreover, $ $ $ground(Rew(\PS)[2])$ contains the preference $A\succ A'$.  
	Therefore, $K \sqsupset M$ and $M$ is not a preferred stable model of $Rew(\PS)$. This is a contradiction.
	
	\item
	($L[\overline{\MP}]= N[\overline{\MP}]\ \wedge\ L[\underline{\MP}]\subset N[\underline{\MP}]$).
	In this case, the maximal mapping rules that will be deleted from $ground(\PS)$ to derive $\PS^L$ 
	coincide with those that will be
	deleted to derive $\PS^N$. Moreover, 
	the ground minimal mapping rules that will be deleted from $ground(\PS)$ to derive $\PS^L$ are a superset of those that will be deleted to to derive $\PS^N$.
	
	
	Let us consider the set $K=L\cup \{A'\ |\ A\not\in L\ \wedge\ (A\leftharpoonup \B\in ground(\PS)\ \vee\ A\leftharpoondown \B\in ground(\PS))\ \wedge\ L\models \B\}$. By construction, it is the minimal model of $Rew(\PS)[1]^{K}$. Then $K$ is a stable model of $Rew(\PS)[1]$. 
	Observe that, must exist two 
	atoms $A'\in K$ and $A\in M$ and, by construction, there cannot exist two atoms $B'\in M$ and $B\in K$. Moreover, $ground(Rew(\PS)[2])$ contains the preference $A'\succ A$.  Therefore, $K \sqsupset M$ and $M$ is not a preferred stable model of $Rew(\PS)$. This is a contradiction.
\end{itemize}

\item ($PSM(\PS)\supseteq \PWM(\PS)$)

Let \ \ $N\in \PWM(\PS)$ \ and \ \ $M=N\cup \{A'\ |\ A\not\in N\ \wedge\ (A\leftharpoonup \B\in ground(\PS)\ \vee\ A\leftharpoondown \B\in ground(\PS))\ \wedge\ N\models \B\}$. First we prove that $M$ is a stable model of $Rew(\PS)[1]$ i.e. it is the minimal model
of $(Rew(\PS)[1])^M$.

By \ construction, 1 $(Rew(\PS)[1])^M=St(\PS^{N})\cup\{A'\leftarrow \B\ |\ A\not\in N\ \wedge\ (A\leftharpoonup \B\in ground$ $(\PS)\ \vee\ A\leftharpoondown \B\in ground(\PS))\ \wedge\ N\models \B\}$.
We have that:
\begin{itemize}
	\item 
	The minimal model of $St(\PS)^{N}$ is $N$, as $N$ is a weak model of $\PS$;
	\item 
	The minimal model of $\{A'\leftarrow \B\ |\ A\not\in N\ \wedge\ (A\leftharpoonup \B\in ground(\PS)\ \vee\ A\leftharpoondown \B\in ground(\PS))\ \wedge\ N\models \B\}$ is 
	$M\setminus N$;
	\item
	Non primed atoms $A$ can be only inferred by rules in $St(\PS^{N})$ and 
	\item
	No primed atom $A'$ occurs in the body of any rule of $St(\PS^{N})$. 
\end{itemize} 
Therefore, \ the \ minimal \ model of $(Rew(\PS)[1])^M$ is $M$ and $M$ is a stable model of $Rew(\PS)[1]$.\\ 

Now we prove that $M$ is a preferred stable model for $Rew(\PS)$. Let us assume by contradiction that there is a stable model $L$ for $Rew(\PS)$ s.t. $L\sqsupset M$. From point 1. and preferences in $Rew(\PS)[2]$, we have that 
$St(L)$ is a weak model for $\PS$ and $St(L)[\overline{\MP}]\supset St(M)[\overline{\MP}]$ or $St(L)[\overline{\MP}]=St(M)[\overline{\MP}] \wedge St(L)[\underline{\MP}]\subset St(M)[\underline{\MP}]$, that is $St(M)$ is not a max-min weak model for $\PS$.
This is a contradiction.~\hfill $\Box$	
\end{enumerate}

\chn{
	This characterization of the Max-Min Weak Model Semantics
	provides a computational mechanism allowing to
	derive the max-min weak models of a P2P system.
}

\chg{
\section{Query Answers and Complexity}\label{compl}
We  consider now the computational complexity of calculating max-min weak models
and answers to queries \cite{DBLP:books/daglib/0072413}.
As a P2P system may admit more than one max-min weak
model, the answer to a query is given by considering \emph{brave} or
\emph{cautious} reasoning (also known as \emph{possible} and \emph{certain} semantics). Issues related to the distributed computation will be discussed in Section \ref{Prototype}.
\vspace{1mm}
\noindent
\begin{definition}
Given a P2P system $\PS$ and a ground peer atom $A$,
then $A$ is \emph{true} under
\begin{itemize}
	\item
	brave reasoning if
	$A \in \bigcup_{M \in MaxMinWM(PS)} M$,
	\item
	cautious reasoning if
	$A \in \bigcap_{M \in  MaxMinWM(PS)} M$.
	~\hfill $\Box$
\end{itemize}
\end{definition}
We assume here a simplified framework not considering the
distributed complexity as we suppose that the complexity of
communications depends on the number of computed atoms which are
the only elements exported by peers. 
\begin{theorem}\label{theorem-PWM1}
Let $\PS$ be a P2P system, then:
\begin{enumerate}
	\item
	Deciding whether an interpretation $M$ is a max-min weak model of $\PS$ is 
	$co\NP$ complete.
	\item
	Deciding whether a max-min weak model for $\PS$ exists is in $\Sigma_2^p$.
	\item
	Deciding whether an atom $A$ is \emph{true} in some max-min weak model of $\PS$ is
	$\Sigma_2^p$ complete.
	\item
	Deciding whether an atom $A$ is \emph{true} in every max-min weak model of
	$\PS$ is  $\Pi_2^p$ complete.
\end{enumerate}
\end{theorem}
\vspace*{-2mm}
\noindent
\textbf{Proof} 
\begin{enumerate}
\item
\textbf{(Membership)}
We prove that the complementary problem, that is the problem of
checking whether $M$ is not a max-min weak model, is in
$\NP$. We can guess an interpretation $N$ and verify in polynomial time that
(i) $N$ is a weak model, that is $\{N\}=\MM(St(\PS^N))$, and
(ii) either $M$ is not a weak model, that is $\{M\}\neq\MM(St(\PS^M))$, or $ N \sqsupset M$, that is $N[\overline{\MP}]\supset M[\overline{\MP}]$ or $N[\overline{\MP}]=M[\overline{\MP}]\wedge N[\underline{\MP}]\subset M[\underline{\MP}]$.
Therefore, the original problem is in $co\NP$. \\
\textbf{(Hardness)}
We will reduce the \emph{SAT problem} to the problem of \emph{checking whether a weak model is not max-min}.
Let $X$ be a set of variables and $F$ a CNF formula over $X$. 
Then the problem that will be reduced is checking whether the QBF formula $(\exists X)\ F $ is $true$.
We define a P2P system $\PS$ with two peers: $\PP_1$ and $\PP_2$.
Peer $\PP_1$ contains the atoms:
\[
\begin{array}{ll}
1\:variable(x) \mbox{, for each $x\in X$}\\
1\:truthValue(true)\\
1\:truthValue(false)
\end{array}
\]
The relation $1\:variable$ stores the variables in $X$ and
the relation $1\:truthValue$ stores the truth values $true$ and $false$.\\
Peer $\PP_2$ contains the atoms:
\[
\begin{array}{ll}
2\:variable(x) \mbox{, for each $x\in X$}\\
2\:positive(x,c)\mbox{, for each $x\in X$ and clause $c$ in F s.t. $x$ occurs non-negated in $c$}\\
2\:negated(x,c)\mbox{, for each $x\in X$ and clause $c$ in F s.t. $x$ occurs negated in $c$}
\end{array}
\]
the mapping rule:
\[
\begin{array}{ll}
2\:assign(X,V)  \leftharpoonup 1\:variable(X), 1\:truthValue(V)
\end{array}
\]
stating that the truth value $V$ could be assigned to the variable $X$, \\
the standard rules:
\[
\begin{array}{ll}
2\:clause(C) \leftarrow 2\:positive(X,C) \\
2\:clause(C) \leftarrow 2\:negated(X,C) \\
2\:holds(C) \leftarrow 2\:positive(X,C),2\:assign(X,true) \\
2\:holds(C) \leftarrow 2\:negated(X,C),2\:assign(X,false) \\
2\:assignment \leftarrow 2\:assign(X,V)\\
\end{array}
\]
defining a $clause$ from the occurrences of its positive and negated variables (first and second rule), whether a clause holds with a given assignment of values (third and fourth rule) and whether an assignment of values actually exists (fifth rule), 
and the integrity constraints:
\[
\begin{array}{ll}
\leftarrow 2\:assign(X,true),\ 2\:assign(X,false) \\
\leftarrow 2\:clause(C),\ not\ 2\:holds(C),\ 2\:assignment\\
\leftarrow 2\:variable(X),\ not\ 2\:assign(X,true),\ not\ 2\:assign(X,false),\ 2\:assignment\\
\end{array}
\]
stating that two different truth values cannot be 
assigned to the same variable (first constraint), that
if there is an assignment then there cannot be an unsatisfied clause (second constraint) and cannot be an unevaluated variable (third constraint).
Let $\D$ the set of atoms in $\PS$, 
$\MP$ the set of mapping rules in $\PS$, $\LP$ the set of standard rules  in $\PS$ and $\IC$ the set of integrity constraints in $\PS$.
Let $M$ be the minimal model of $\D\cup\LP\cup\IC$, that is the model containing no mapping atom. 
As $\PS$ is locally consistent, $M$ is a weak model of $\PS$.
Observe that, the integrity constraints in $\PS$ are satisfied when no mapping atom is imported in $\PP_2$ that is if no assignment of values is performed for the variables in $X$.
If $F$ is not satisfiable, then there is no way to import mapping atoms in $P_2$ preserving consistency because the second constraint will be violated. In this case $M$ is a max-min weak model. If $F$ is satisfiable there is a weak model $N$ whose set of mapping atoms corresponds to an assignment of values to the variables in $X$ that satisfies $F$.
Clearly, as $\MP[N]\supset\MP[M]$, $M$ is not a max-min weak model. Moreover, if $M$ is not a max-min weak model there must be another weak model $N$ whose set of mapping atoms corresponds to an assignment of values to the variables in $X$ that 
satisfies $F$.
In other words, $F$ is satisfiable if and only if $M$ is not a max-min weak model. \\
\item
Let us guess an interpretation $M$. By (1), deciding whether $M$ is a max-min weak model
can be decided by a call to a $co\NP$ oracle. \\
\item
From Theorem \ref{equivalence}, an atom $A$ is $true$ in some
max-min weak model of $\PS$ if and only if it is true in
some preferred stable model of $\PS$. 
The complexity of this problem has been presented in \cite{SakIno00}. For disjunction-free ($\vee-free$)
prioritized logic programs, deciding
whether an atom is \emph{true} in some preferred stable model is
$\Sigma_2^p$ complete.\\
\item
From Theorem \ref{equivalence}, an atom $A$ is $true$ in every
max-min weak model of $\PS$ if and only if it is true in
every preferred stable model of $\PS$. 
The complexity of this problem has been presented in \cite{SakIno00}. For disjunction-free ($\vee-free$)
prioritized logic programs, deciding
whether an atom is \emph{true} in every preferred stable model is
$\Pi_2^p$ complete.
\hfill $\Box$
\end{enumerate}
}
\ch{
\section{Discussion}\label{Discussion}
This section introduces some useful discussions on some features of the proposed semantics. \\
\subsection{Dealing with Locally Inconsistent P2P Systems}
\ch{
	The framework presented so far does not guarantee that a locally inconsistent P2P system (i.e. containing at least a locally inconsistent peer) has a weak model.
	Indeed, there could be locally inconsistent peers that cannot reach a consistent state by importing sets of atoms from other peers. 
	This happens because the only mechanism modeled by 
	our original framework is an `enriching' mechanism that does not allow deletions of atoms from local databases. 
	\\
	Therefore, if a peer is locally inconsistent and there is no way to
	import mapping atoms able to restore its consistency, the peer remains inconsistent because no atom can be deleted.
	\\
	In this section, we present an extension of our framework that 
	\textit{simulates} deletions of atoms by using maximal mapping rules.\\
	Informally, the idea is to create for each peer an \textit{auxiliary peer}, move the database from the original peer to the auxiliary one and equip the original peer with a set of
	maximal mapping rules allowing to import, from the auxiliary peer, maximal sets of atoms 
	not violating its integrity constraints.
	\begin{definition}
		Let $\PS=\{\PP_1,\dots,\PP_n\}$  and $\PP_i=\<\D_i, \LP_i, \MP_i, \IC_i\>$, with $i\in[1..n]$, a peer in $\PS$.\\
		Then, $Split(\PP_i)$ is the set containing the following peers:
		\chn{
		\begin{itemize}		
			\item $\PP_{(i+n)}'=\<\{(i+n):p(X)\ |\ i:p(X)\in\D_i\}, \emptyset, \emptyset, \emptyset\>$ \\
			\item $\PP_{i}'=\<\emptyset, \LP_i, \MP_i\cup\ \widehat{\MP_i}, \IC_i\>$, where  $\widehat{\MP_i}=\{i:p(X)\leftharpoonup (i+n):p(X)\ |\ i:p(X)\in\D_i\}$
		\end{itemize}
		}
		Moreover:
		$$Split(\PS)=\bigcup_{\PP_i\ \in\ \PS} Split(\PP_i)$$
		\hfill $\Box$
	\end{definition}
\chn{
	In the previous definition, peer $\PP_i'$ is derived from peer $\PP_i$ by deleting its local database $\D_i$ and inserting a set of maximal mapping rules allowing to import facts into the old base relations (which now are mapping relations) from the auxiliary peer $\PP_{(i+n)}'$.
}
	Given a P2P system $\PS$, we define $\widehat{\MP}=\bigcup_{P_i\in PS} \widehat{\MP_i}$.\\
	We now present a generalization of our semantics allowing to deal with locally inconsistent P2P systems.
	\begin{definition}
		Let $\PS=\{\PP_1,\dots,\PP_n\}$. The \textit{generalized weak models} of $\PS$, denoted as $GWM(\PS)$, are obtained from the weak models of $Split(\PS)$ by removing all the atoms $i\:A$ with $i>n$.
	\end{definition}
	\begin{definition}
		\textsc{[Generalized Max-Min Weak Model]} 
		\ch{Given a P2P system $\PS$} and two generalized weak models $M$ and $N$ \ch{of $\PS$}, we say that $M$ is \emph{G-preferable}
		to $N$, and we write $M \sqsupseteq_G N$, if
		\begin{itemize}
			\item
			$M[\widehat{\MP}] \supset N[\widehat{\MP}]$ \ \ or
			\item
			$M[\widehat{\MP}] = N[\widehat{\MP}]$ and
			$M[\overline{\MP}] \supset N[\overline{\MP}]$ \ \ or
			\item
			$M[\widehat{\MP}] = N[\widehat{\MP}]$ and $M[\overline{\MP}] = N[\overline{\MP}]$ and
			$M[\underline{\MP}] \subseteq N[\underline{\MP}]$
		\end{itemize}
		Moreover, if $M \sqsupseteq_G N$ and $N \not\sqsupseteq_G M$ we write $M
		\sqsupset_G N$.
		A weak model $M$ is said to be a \textit{generalized Max-Min weak model} if there is no weak model
		$N$ such that $N \sqsubset_G M$. The set of generalized Max-Min weak models  will be
		denoted by $GMinMaxWM(\PS)$.
		~\hfill $\Box$
	\end{definition}
	\begin{example}\label{ExInconsistent}
		\begin{figure}[h]
			\centering
			\includegraphics[width=0.8\textwidth]{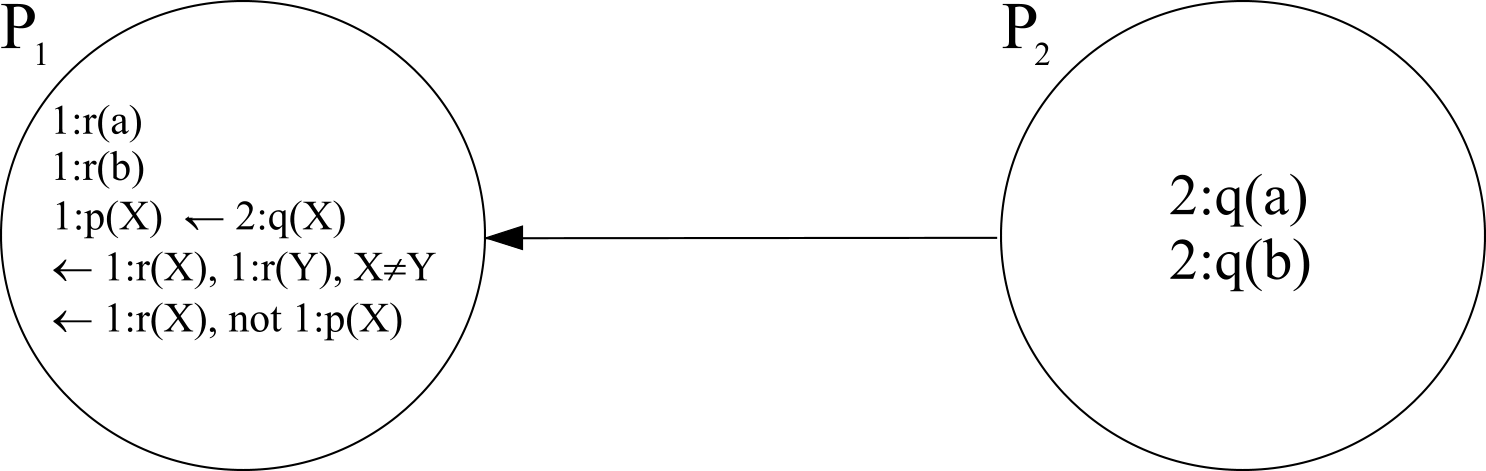}
			\caption{The system $\PS$}
			\label{figInconsistent}
		\end{figure}
		\noindent Consider the P2P system $\PS$ depicted in Figure
		\ref{figInconsistent}. $\PP_2$ contains the facts $2\:q(a)$ and $2\:q(b)$, whereas
		$\PP_1$ contains
		the minimal mapping rule $1\:p(X) \leftharpoondown 2\:q(X)$;
		the  constraint $\leftarrow 1\:r(X), 1\:r(Y), $ $X\!\neq\!Y$ stating that the base relation $1\:r$ can contain at most one tuple, the  constraint $\leftarrow 1\:r(X),not\ $ $1\:p(X)$ stating that if $\PP_1$ contains the fact $1\:r(X)$ then the fact $1\:p(X)$ has to be derived; and the facts $1\:r(a)$ and $1\:r(b)$.\\
		This P2P system is inconsistent because the local database of peer
		$\PP_1$ violates the constraint $\leftarrow 1\:r(X), 1\:r(Y), X\!\neq\!Y$. \\
		$Split(\PS)$ is depicted in Figure \ref{figSplit}.\\
		\begin{figure}[h]
			\centering
			\includegraphics[width=0.8\textwidth]{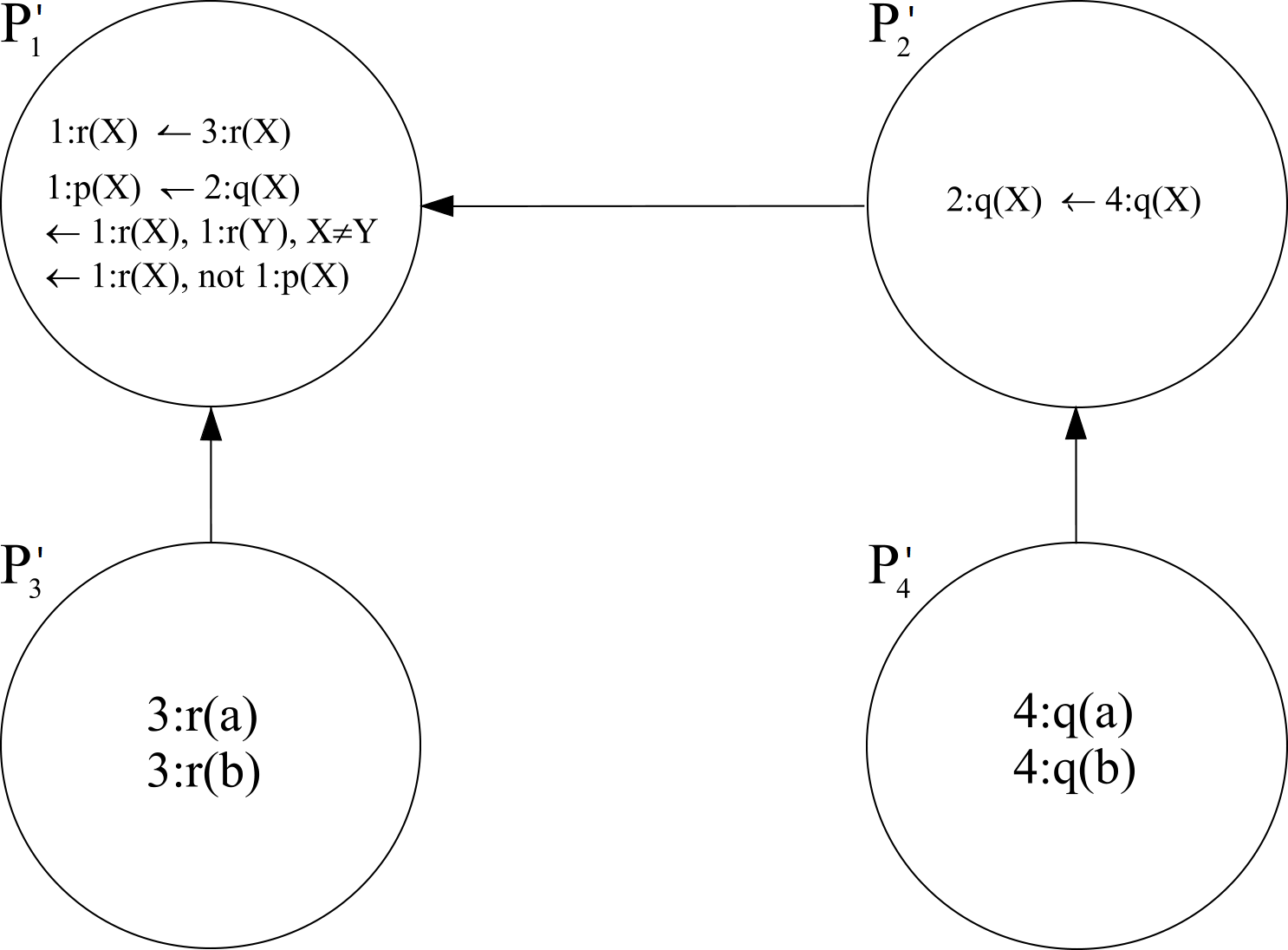}
			\caption{The system $Split(\PS)$}
			\label{figSplit}
		\end{figure}
		\noindent
		The generalized Max-Min weak models of $\PS$ are: 
		\begin{itemize}
			\item 
			$M_1=\{2\:q(a),2\:q(b),1\:r(a),1\:p(a)\}$ 
			\item 
			$M_2=\{2\:q(a),2\:q(b),1\:r(b),1\:p(b)\}$.
		\end{itemize}
		~\hfill $\Box$
	\end{example}
}
\chn{
Observe that if $LP$ contains negation this
technique does not guarantee that consistency 
can be achieved.
}
\subsection{Computing the Max Weak Model Semantics}\label{Rew}
This section recalls an alternative characterization of the
max weak model semantics that allows to model a maximal P2P system $\PS=\<\D,\LP,\MP,\IC\>$, where $\LP$ only contains \textit{positive} peer standard rules, with a single disjunctive logic program $Rew_t(\PS)$ embedding the set of priorities presented in Section \ref{RewMaxMin} \cite{SUM2007}.\\
In \cite{SUM2007} it has been shown that the stable models of $Rew_t(\PS)$ correspond to the maximal weak models of $\PS$.\\
Using this rewriting, the computation of the preferred
weak models of a P2P system $\PS$ can be performed in a \emph{centralized way} by using an inference engine like DLV \cite{DBLP:conf/sigmod/LeoneGILTEFFGRLLRKNS05} able to process
$Rew_t(\PS)$ and compute its stable models.\\
Although this approach is still not pragmatic, because the computation
is centralized and its complexity is prohibitive for real cases,
 the program $Rew_t(\PS)$ can be used as a starting
point for a \emph{distributed technique}, as will be pointed out in 
Section \ref{Prototype}.\\
The formal details of this approach are out of the scope of this
section and can be found in \cite{SUM2007}.
Here we want to show how to make our 
approach pragmatic by implementing a derived version of max weak model
semantics into a real P2P system.
\\
Let's firstly introduce some concepts and definitions.
Given a peer atom $A=i\:p(x)$, $A^t$  denotes the atom $i\:p^t(x)$ and  $A^v$ denotes the atom $i\:p^v(x)$.
$A^t$ is called \emph{testing atom}, whereas $A^v$ is called  \emph{violating atom}.\\
\chn{
A testing atom $A^t$ corresponds to the mapping atom $A$ that could be derived in the target peer. 
While $A$ is derived only if its existence does not cause any inconsistency, 
$A^t$ is always derived in order to test whether $A$ can be inferred safely.
If the presence of $A$ violates at least one
integrity constraint, the corresponding violating atom $A^v$ is derived. 
In this case, the atom $A^v$ blocks the derivation of $A$ and the inconsistency that $A$ would cause is prevented. 
}
\begin{definition}
	\label{conj}
	Given a conjunction
	\begin{equation}\label{body-conj}
	\hspace{-1mm}  \B=A_1,\dots,A_h,not\ A_{h+1},\dots,not\ A_n,B_1,\dots,B_k,not\ B_{k+1},\dots,not\ B_m,\varphi
	\end{equation}
	where $A_i$ ($i\in[1..\ n]$) is a mapping atom or a derived atom, $B_i$ ($i\in[1..\ m]$) is a base atom and
	$\varphi$ is a conjunction of built in atoms, we define
	\begin{equation}\label{Rew-body-conj}
	\hspace{-1mm} \B^t=A^t_1,\dots,A^t_h,not\ A^t_{h+1},\dots,not\ A^t_n,B_1,\dots,B_k,not\ B_{k+1},\dots,not\ B_m,\varphi
	\end{equation}
	~\hfill $\Box$
\end{definition}
From the previous definition it  follows that  given  a negation free conjunction of the form
\begin{equation}\label{NFbody-conj}
\B=A_1,\dots,A_h,B_1,\dots, B_k,\dots,\varphi
\end{equation}
\noindent
then
\begin{equation}\label{RewNF-body-conj}
\B^t=A^t_1,\dots,A^t_h,B_1,\dots,B_k,\varphi.
\end{equation}
\begin{definition}
	\textsc{[Rewriting of an integrity constraint].} \ 
	Given  \ an  \ integrity constraint\footnote{\ch{Recall that $\B$ is of the form (\ref{body-conj}).}}
	$
	c  =\ \ \ \leftarrow \B
	$,
	its \ rewriting \ is \ defined \ as \
	$Rew_t(c)=\{A_1^v \vee \dots \vee A_h^v \leftarrow \B^t \}$.~\hfill $\Box$
\end{definition}
\noindent
If the body $\B^t$ (that is of the form (\ref{Rew-body-conj})),
in the previous definition is \emph{true},
then it can be deduced  that at least one of the
violating atoms $A_1^v,\dots,A_h^v$ is \emph{true}.
This states that in order to avoid  inconsistencies,
at least one of the atoms $A_1,\dots,A_h$ cannot be inferred.
\vspace{2mm}
\begin{definition}
	\textsc{[Rewriting of a standard rule].} \ 
	Given a standard rule\footnotemark[4]
	$
	s = H \leftarrow \B
	$,
	its rewriting is defined as
	$Rew_t(s)=\{H \ \leftarrow \B;\ H^t \leftarrow \B^t;\ A_1^v \vee \dots \vee A_h^v \leftarrow \B^t,H^v\ \}$.
	~\hfill $\Box$
\end{definition}
In order to find the mapping atoms that, if imported, generate some inconsistencies
(i.e. in order to find their corresponding violating atoms), all possible
mapping testing atoms are imported and the derived testing atoms are inferred.
\noindent
In the previous definition, if $\B^t$ (that  is of the form (\ref{RewNF-body-conj})),
is \emph{true} and the violating atom $H^v$ is \emph{true}, then the body of the disjunctive rule
is \emph{true} and therefore it can be deduced
that
at least one of the
violating atoms $A_1^v,\dots,A_h^v$ is \emph{true} (i.e. to avoid such inconsistencies
at least one of atoms $A_1,\dots,A_h$ cannot be inferred).\\
\begin{definition}
	\textsc{[Rewriting of a maximal mapping rule]} 
	Given a mapping rule\footnote{\ch{Recall that $\B$ is of the form (\ref{NFbody-conj}).}}
	$
	m = H  \leftharpoonup \B
	$, its rewriting is defined as
	$Rew_t(m)=\{H^t \leftarrow \B;\ H\  \leftarrow H^t, not \ H^v\ \}$.~\hfill $\Box$
\end{definition}
Intuitively, to check whether a mapping atom $H$ generates some inconsistencies,
if imported in its target peer, a testing atom $H^t$ is imported in the same
peer.
Rather than violating some integrity constraint, it (eventually) generates, by rules obtained from the rewriting of standard rules and integrity 
constraints, the atom $H^v$.
In this case $H$,  cannot be inferred and inconsistencies are prevented.
\begin{definition}
	\label{rewp2p}
	\textsc{[Rewriting of a Maximal P2P system]} \ 
	Given a Maximal P2P system $\PS = \D \cup \LP \cup \MP \cup \IC$, then
	\begin{itemize}
		\item $Rew_t(\MP)=\bigcup_{m \in {\cal MP}}Rew_t(m)$      \vspace{1mm}
		\item $Rew_t(\LP)=\ \bigcup_{s \in {\cal LP}}Rew_t(s)$    \vspace{1mm}
		\item $Rew_t(\IC)=\ \ \bigcup_{c \in {\cal IC}}Rew_t(c)$  \vspace{1mm}
		\item $Rew_t(\PS)=\D \cup Rew_t(\LP) \cup Rew_t(\MP) \cup Rew_t(\IC)$
		~\hfill $\Box$
	\end{itemize}
\end{definition}
\vspace{3mm}
\begin{example}\label{example-max}
	Let us consider the maximal P2P system presented in Example \ref{Ex3-maximal}.From Definition  (\ref{rewp2p}) we obtain:\\
	\begin{tabular}{ll}
		\noindent
		$Rew_t(\PS)=$&\hspace{-1mm}$\{2\:q(a);\ 2\:q(b);$ \\
		& \hspace{-0.5mm} $ 1\:p^t(X)\leftarrow 2\:q(X);$ \\
		& \hspace{-0.5mm} $  1\:p(X)\leftarrow 1\:p^t(X),not\ 1\:p^v(X)$;\\
		&\hspace{-0.5mm}  $1\:p^v(X) \vee 1\:p^v(Y)\leftarrow 1\:p^t(X),1\:p^t(Y),X \neq Y\}$
	\end{tabular}
 	\vspace{2mm}
	\noindent
	The stable models of $Rew_t(\PS)$ are:
	\newline
	\begin{tabular}{l}
		$M_1=\{2\:q(a),2\:q(b),1\:p^t(a),1\:p^t(b),1\:p^v(a),1\:p(b)\},$\\
		$M_2=\{2\:q(a),2\:q(b),1\:p^t(a),1\:p^t(b),1\:p(a),1\:p^v(b)\}$
	\end{tabular}
\hspace*{8cm}\hfill $\Box$
\end{example}
\begin{definition}
	\textsc{[Total Stable Model]} \ 
	Given a P2P system $\PS$ and a stable model $M$ for $Rew_t(\PS)$,
	the interpretation obtained by deleting from $M$ its violating and testing atoms, denoted as
	$\T(M)$,
	is a \emph{total stable model} of $\PS$.
	The set of total stable models of $\PS$ is denoted as $\TSM(\PS)$.
	~\hfill $\Box$
\end{definition}
\begin{example}\label{tsmExample-cont}
	For the P2P system $\PS$ reported in Example \ref{example-max},
	$\TSM(\PS)=\{\{2\:q(a),2\:q(b),1\:p(b)\},\{2\:q(a),$ $2\:q(b),1\:p(a)\}\}$.
	~\hfill $\Box$
\end{example}
In \cite{SUM2007} it has been shown that the set of total stable models is equivalent to the set of maximal weak models, i.e. $\TSM(\PS)=MaxWM(\PS)$.
\\
Observe that,  this rewriting technique allows computing
the maximal weak models of a P2P system  with an arbitrary topology.
The topology of the system will be encoded in  its rewriting.
As an example, if a system $\PS$ is \emph{cyclic}, its rewriting
$Rew_t(\PS)$ could be \emph{recursive}.
\subsection{A System Prototype}
\label{Prototype}
The rewriting presented in the previous section has been used in 
\cite{SUM2007,adbis2017,iiwas2017}
as a starting point to implement a system prototype of a P2P
system based on a deterministic version of our maximal weak model
semantics.
\\
The first important observation is that a P2P system may admit many maximal weak models whose computational complexity has been shown to be prohibitive.
\\
Therefore, it is needed to look for
a more  pragmatic solution for assigning a semantics to a P2P system. Starting from this observation, a deterministic
model whose computation is guaranteed to be polynomial time  has been proposed in \cite{SUM2007,adbis2017,iiwas2017}. 
The new proposed semantics, called \textit{well founded semantics}, assigns to a P2P system its \emph{Well Founded Model}, a three valued partial deterministic model that  captures the intuition that if an atom is \textit{true} in a maximal weak model and it is \textit{false} in another one, 
then it is \textit{undefined} in the well founded model \cite{DBLP:conf/pods/Gelder89,DBLP:conf/cl/LoncT00}.
\\
It has been shown that, given a maximal P2P system $\PS$ whose standard rules are positive, the rewriting $Rew_t(\PS )$ presented in Section \ref{Rew} is Head Cycle Free \cite{DBLP:conf/iclp/Ben-EliyahuD92}. 
Therefore, it can be normalized obtaining a normal program that we denote as $Rew_w(\PS)$. 
\\
The next step is to adopt for $Rew_w(\PS)$ a Well Founded Model
Semantics. The program, $Rew_w(\PS)$
admits a well founded model $W$ that can be computed in polynomial time.
\begin{example}\label{ex-WF}
	Consider the P2P system presented in Example \ref{Ex3-maximal}.
	The normal version $Rew_w(\PS)$ of the rewriting $Rew_t(\PS)$ presented in Example \ref{example-max} is:\\ \\
		\begin{tabular}{ll}
		\noindent
		$Rew_w(\PS)=$&\hspace{-1mm}$\{2\:q(a);\ 2\:q(b);$ \\
		& \hspace{-0.5mm} $ 1\:p^t(X)\leftarrow 2\:q(X);$ \\
		& \hspace{-0.5mm} $  1\:p(X)\leftarrow 1\:p^t(X),not\ 1\:p^v(X)$;\\
		&\hspace{-0.5mm}  $1\:p^v(X) \leftarrow 1\:p^t(X),1\:p^t(Y),X \neq Y\}$\\
		&\hspace{-0.5mm}  $1\:p^v(Y)\leftarrow 1\:p^t(X),1\:p^t(Y),X \neq Y\}$\\  \\
	\end{tabular}
	The well founded semantics of $\PS$ is given by the well founded model of $Rew_w(\PS)$, $W=\<\{2\:q(a),2\:q(b)\},\emptyset\>$\footnote{\ch{The first component of the pair is the set of \textit{true} facts while the second one is the set of \textit{false} facts.}}.
	The facts $2\:q(a)$ and $2\:q(b)$ are \emph{true}, while the facts
	$1\:p(a)$ and $1\:p(b)$ are \emph{undefined}.
	\hfill $\Box$
\end{example}
Although the adoption of a well founded model for a maximal P2P system represents a step forward in the implementation of a real system prototype -- as it can be computed in polynomial time--
 it is evident that the evaluation of a unique logic program requires  a \emph{centralized computation} and this is not 
 realistic: a \emph{distributed computation} is needed. \\ 
In \cite{adbis2017,iiwas2017}  a technique allowing to compute the well founded model in a distributed way has been presented.\\
The basic idea is that each peer computes its own portion of $Rew_w(\PS)$, sending the result to the other peers.
In more detail, if a peer receives a query, then it recursively queries  the peers to which 
it is connected through mapping rules, before being able to calculate its answer. 
\\
Formally, a local query submitted by a user to a peer does not differ from a remote query submitted by another peer. 
The only substantial difference is in the construction of the answer that in the case of remote query, must be returned to the requesting peer. 
Once retrieved the necessary data from neighbor peers, the peer computes its  well founded model and evaluates the query (either local or remote) on that model. If the query is a remote query, the answer is sent to the requesting peer.\\
Details  on the architecture  and  implementation of 
this system prototype can be found in \cite{iiwas2017}. 
The paper in \cite{iiwas2017} also reports an application scenario related to the integration of biomedical data from PubMed  (http://www.ncbi.nlm.\\
nih.gov/pubmed/).
The experiment has been conducted by considering three peers: \emph{Peer1}, \emph{Peer2} and \emph{Peer3}. \emph{Peer1} contains information about papers related to the HIV virus, \emph{Peer2} about papers related to the Ebola virus and \emph{Peer3} integrates data provided by \emph{Peer1} and \emph{Peer2}. The final  aim of this experiment is the integration of all the papers related to both HIV and Ebola virus into a unique data source in \emph{Peer3}.
}

\section{Related Works}\label{RW}

\paragraph{\bf Semantic Peer Data Management Systems.}

\nop{-------------------------
	\noindent
	In \cite{TatHal04}  several techniques for optimizing the
	reformulation of queries in a PDMS are presented. In particular the
	paper presents techniques for pruning semantic paths of mappings  in
	the reformulation process and for minimizing the reformulated
	queries.
	
	\noindent
	The design of optimization methods for query processing
	over a network of semantically related data  is investigated in
	\cite{MadAhl*03}.
	------------------------------------------------------------}

The present paper is placed among the works on \emph{semantic peer data management systems}. 
\ch{
	This research topic formally started with the work in \cite{Halevy*03} in which 
	 the problem of schema mediation in a P2P system is investigated
	A  formalism,
	$PPL$, for mediating peer schemas, which uses the GAV and LAV \cite{DBLP:conf/pods/Lenzerini02}
	formalism to specify mappings, is proposed.
	Mappings relate two conjunctive queries expressed in terms of the schema of disjoint peers.
	The semantics is assigned   using classical first-order logic (FOL) and query
	answering is defined by extending the notion of certain
	answer.
 More specifically, certain answers for a peer are those that are true in every global instance 
 that is consistent with local data. This choice implies, as a consequence, the need for   
	the consistency of each peer 
	with respect to the whole P2P system.
	 As for a comparison, in this paper we do not adopt a FOL interpretation for a P2P system and tolerate inconsistencies.
	A mapping  rule \cite{Halevy*03} is a logical implication between two peers. It is often the case that  preference
	is given to external data over internal data, or equivalently by using the concept of \emph{trust} given in
	\cite{DBLP:journals/tplp/BertossiB17}  $(P, less, Q)$ denotes that the peer $P$ trusts itself 
	less than $Q$. This paper formalizes a different  proposal: mapping rules are a means used
	 either to import maximal sets of
	atoms  while
	preventing inconsistency anomalies  or to fix the knowledge by importing the 
	minimal sets of atoms allowing to restore consistency. 
	In both cases, in our basic framework, we implicitly satisfy the preference that a peer trusts more its own data over data provided by other peers.}
\noindent

\noindent
In \cite{Calv*04} a sound, complete and terminating
procedure that returns the certain answers to a query submitted to a
peer, is proposed. The paper presents 
a  semantics for a P2P system, based
on epistemic logic. 
\ch{Mapping rules between two peers $\PP_1$ and $\PP_2$ are of the form
	 $CQ_1 \rightarrow CQ_2$, where $CQ_1$ and $CQ_2$ are conjunctive queries over the schema of $\PP_1$ and $\PP_2$.
 }
An advantage of this framework is that certain answers of fixed conjunctive queries posed 
on a peer can be computed in polynomial time. 
\ch{ The proposal does not manage local inconsistency. Each peer has to be consistent with  respect to its integrity constraints, otherwise the entire P2P system is considered inconsistent. Moreover, if inconsistencies arise due to mapping rules the whole P2P system is considered  inconsistent.
	An extension of the epistemic theory that ensures \emph{local inconsistency tolerance} has been 
	presented in \cite{DBLP:journals/is/CalvaneseGLLR08}. The paper extends the epistemic theory with an additional operator so as to  tolerate local inconsistency. More specifically, it ignores a peer inconsistent with respect to its own local constraints. No consistency restoration process is proposed; on the contrary in our proposal an inconsistent peer is not cut off of the system, but can be \textit{repaired} by means of 
	mapping rules so as to restore consistency.  
}

\noindent
In \cite{Franconi*04a,DBLP:conf/vldb/FranconiKLZ04,DBLP:conf/edbtw/FranconiKLZ04} a characterization of P2P database
systems and a model-theoretic semantics dealing with inconsistent
peers is proposed. The basic idea is that if a peer does not have
models all the (ground) queries submitted to the peer are \emph{true}
(i.e. are \emph{true} with respect to all models). Thus, if some
databases are inconsistent it does not mean that the entire system
is inconsistent.
The semantics in \cite{Franconi*04a} coincides with
the epistemic semantics in \cite{Calv*04,CalvaneseGLLR05}.
\ch{The semantics in \cite{Franconi*04a,DBLP:conf/vldb/FranconiKLZ04,DBLP:conf/edbtw/FranconiKLZ04} 
	also provides a distributed algorithm to compute queries; the setting assumes the existence of a super peer instructor that updates peers' data. The proposal is not inconsistency tolerant as the arise of an inconsistency causes the entire P2P system to became inconsistent.
}
\ch{
As for a comparison with the present proposal, the works in \cite{Calv*04,CalvaneseGLLR05,DBLP:journals/is/CalvaneseGLLR08}
and in \cite{Franconi*04a,DBLP:conf/vldb/FranconiKLZ04,DBLP:conf/edbtw/FranconiKLZ04} 
are significantly different.\\
Consider the P2P system in Fig. \ref{fig}.
\begin{figure}[h]
	\centering
	\includegraphics[width=0.9\textwidth]{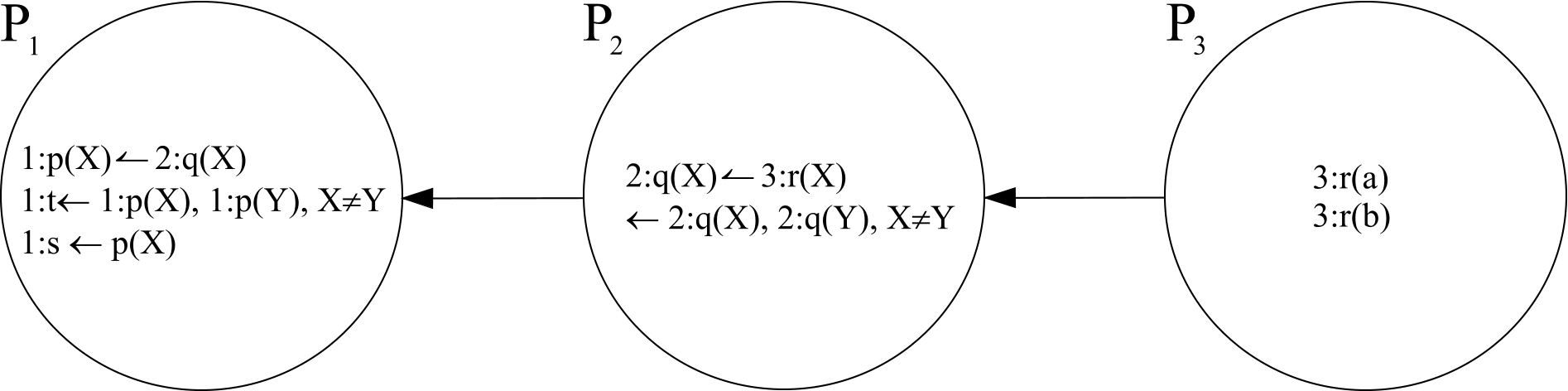}
	\caption{A P2P system}\label{fig}
\end{figure}
The epistemic semantics proposed in \cite{Calv*04,DBLP:journals/is/CalvaneseGLLR08} states that both
atoms $2\:q(a)$ and $2\:q(b)$ are imported in peer $\PP_2$ which
becomes inconsistent. In this case the semantics assumes that the
whole P2P system is inconsistent and every atom is \emph{true} as it
belongs to all minimal models. Consequently, $1\:t$ and $1\:s$
are \emph{true}.
The semantics proposed in \cite{Franconi*04a}
assumes that only $\PP_2$ is inconsistent as it has no model. Thus,
as the atoms $2\:q(a)$ and $2\:q(b)$ are \emph{true} in $\PP_2$
(they belong to all  models of $\PP_2$) and the atoms
$1\:p(a)$ and $1\:p(b)$ can be derived in $\PP_1$. Finally,
$1\:t$ and $1\:s$ are  \emph{true}.
The Maximal weak model semantics, here proposed, 
states that with $3\:r(a)$ and $3\:r(b)$ being
\emph{true} in $\PP_3$,  either $2\:q(a)$ or $2\:q(b)$ could be imported
in $\PP_2$ (but not both, otherwise the integrity constraint is
violated) and, consequently, only one tuple is imported in the
relation $1\:p$ of the peer $\PP_1$. Note that, whatever is the
derivation in $\PP_2$, $1\:s$ is derived in $\PP_1$ while $1\:t$ is not
derived. Therefore, the atoms $1\:s$ and $1\:t$ are, respectively,
\emph{true} and \emph{false} in $\PP_1$. 
Considering Example \ref{Motivating-Example-minimal}, as the peer $\PP_2$ is inconsistent
the semantics presented in \cite{Calv*04,DBLP:journals/is/CalvaneseGLLR08} cut it off from the system,
whereas our semantics restore consistency in $\PP_2$ by importing 
either  $2\:supplier(dan,$ $ laptop)$
or $2\:supplier(bob, laptop)$.
}

\ch{
In all previous proposals mapping rules are of `import kind'. None of them uses mapping rules to fix the knowledge of a
correct, but incomplete database as we do by means of minimal mapping rules. They all adopt the traditional classical idea of importing knowledge and use mapping rules as logical implications.
In this paper, we follow a different perspective. 
Maximal mapping rules are used by a peer to import as much knowledge as possible, while preventing inconsistencies, whereas 
minimal mapping rules are used by a peer as a means to restore consistency by importing minimal sets of data.
Moreover, the combined use of both maximal and minimal mapping rules 
allows to characterize each peer in the neighborhood as a resource
used either to enrich (integrate) or to fix (repair) the knowledge,
so as to define a kind of \emph{integrate\-repair} strategy.
This feature has no counterpart in the above mentioned proposals.
}

\paragraph{\bf Preferences in P2P Data Management Systems.}

\ch{In a more general perspective, interesting semantics for data exchange systems,  that offer  the possibility of \ch{explicitly} modeling some preference criteria
while performing the data integration process,  has been  proposed
in \cite{BertossiB04,BerBra2007,DBLP:journals/tplp/BertossiB17,ISMIS2008,IDEAS2011}.}
\noindent
In \cite{BertossiB04,BerBra2007,DBLP:journals/tplp/BertossiB17}
a semantics is proposed that allows for cooperation among pairwise peers that are related to each other by means
 of data exchange constraints (i.e. mapping rules) and trust relationships.
The decision by a peer  on what other data to consider (besides its local data) does not
depend only on its data exchange constraints, but also on the trust relationship that it has with
other peers.
Given a peer $P$ in a P2P system
a solution for $P$ is a database instance that respects the exchange constraints and trust
relationship  $P$ has with its `immediate neighbors'.
Trust relationships are of the form: $(P,less, Q)$ stating
that $P$ trusts itself less that $Q$, $(P,more, Q)$ stating
that $P$ trusts itself more that $Q$ and $(P,same, Q)$ stating
that $P$ trusts itself the same as $Q$.  These trust relationships are static and are used in the process of collecting data in order to establish  preferences in the case of conflicting information. 

The introduction of preference criteria among peers is out of the scope of this paper, 
\ch{
and in the present proposal no explicit preference is formally defined. In any case, note that, an implicit preference 
is embedded into maximal and minimal mapping rules. Specifically, maximal mapping rules state that it is preferable to import as long as no local inconsistencies arise; whereas minimal mapping rules state that it is preferable not to import unless a local inconsistency exists.
In addition, a second implicit level of preference exists in our proposal. Each peer trusts more local data over imported data, therefore our framework always gives more preference to local data  over data imported  by external peers.
} 
\ch{This setting can be easily modified in order to cope with the different perspectives in which a generic peer trusts less or the same its own data w.r.t. data provided by external peers.
}

We have proposed in recent papers extensions of the Max Weak Model Semantics that allow to \ch{explicitly} express preferences between peers: in \cite{ISMIS2008}  a mechanism is defined that allows to set different degrees of reliability for neighbor peers. 
\ch{More specifically, the paper extends the Max Weak Model Semantics with a mechanism that allows to set priorities among mapping rules. While collecting data it is quite natural for a source peer to associate different degrees of reliability to the portion of data provided by its neighbor peers. Starting from this simple observation, the paper in \cite{ISMIS2008} enhances the Max Weak Model Semantics by using priority levels among mapping rules in order to select the maximal weak models containing a maximum number of mapping atoms according to their importance. Trusted Weak Models can be computed as stable models of a logic program with weak constraints \cite{DBLP:journals/tkde/BuccafurriLR00,DBLP:journals/fuin/CalimeriFPL06}.
		}
Both in  \cite{ISMIS2008}  and in \cite{BertossiB04,DBLP:journals/tplp/BertossiB17} the mechanism is \emph{rigid} in the sense that the preference among  conflicting sets of atoms that a peer can import from only depends on the priorities (trust relationship) fixed at design time.  
To overcome static preferences, in  \cite{IDEAS2011} `dynamic'  preferences,  that allow to select among different scenarios looking at the properties of data provided by the peers, is introduced.
The work in  \cite{IDEAS2011} allows to model concepts like \emph{``in the case of conflicting information, it is preferable to import data from the neighbor peer that
	can provide the maximum number of tuples"} or \emph{``in the case of conflicting information, it is preferable to import data from the neighbor peer such that the sum of the values of an attribute is minimum"} without selecting a-priori preferred peers. \\

\paragraph{\bf Relationship to  Multi-Context Systems.}
\noindent	General peer to peer data management systems are related to Multi-Context Systems
		(MCS).
			A MCS consists of a set of contexts and a set of inference rules (known as
		mapping or bridge rules) that allows the information flow between different contexts.
		 The  general nonmonotonic MCS model has been defined in \cite{BreEit2007}.  The paper
	    proposes a general framework for multi-context reasoning
		that enables to combine arbitrary monotonic and nonmonotonic
		logics. Information flows among contexts by means of nonmonotonic bridge rules  and 
		several different notions of equilibrium for  acceptable belief have been investigated.

	In \cite{DBLP:conf/kr/EiterFSW10,DBLP:journals/ai/EiterFSW14} inconsistencies are analyzed in MCSs, in order to understand where and why they occur and how they can be managed.
	Each context is assumed to be consistent, therefore the reason of inconsistencies just relies on the application of mapping rules.
	The paper introduces two approaches of explaining inconsistencies in MCSs in terms of bridge rules:
	the first notion  characterizes inconsistencies in terms of mapping rules that need to be altered to restore consistency, and the second notion  looks for combinations of rules which cause inconsistency. The two notions,
	following the classical terminology  in \cite{DBLP:journals/ai/Reiter87}, are called respectively \emph{diagnosis} and \emph{explanation}. 
	
	\chrr{
		Using the concept of diagnosis it is possible to capture the semantics of maximal P2P systems 
		in terms of MCSs. A Multi Context System $M$ is a collection of contexts $(C_1,\dots,C_n)$, where $C_i = (L_i, kb_i, br_i)$, $L_i$ is a \emph{logic}, $kb_i$ is a knowledge base and $br_i$ is a set of \emph{bridge rules}, for $i\in[1..n]$.
		An equilibrium $S$ is a tuple $(S_1,\dots,S_n)$, where, for $i\in [1..n]$, $S_i$ is the knowledge derived from $kb_i$ and the heads of the bridge rules in $br_i$ whose bodies are satisfied by $S$.
		We consider equilibria that are minimal under component wise set inclusion.\footnote{For precise definitions of these concepts see \cite{BreEit2007}}\\
		A diagnosis is a pair $(D_1,D_2)$, where $D_1$ and $D_2$ are subsets of $\bigcup_{i\in[1..n]} br_i$, such that removing from $M$ the bridge rules in $D_1$ and adding to $M$ the bridge rules of $D_2$ in inconditional form (obtained from the rules in $D_2$ by removing the bodies), $M$ is consistent.
		A maximal P2P system $\PS = \{\PP_1,\dots,\PP_n \}$, where
		$\PP_i = \<\D_i, \LP_i, \MP_i, \IC_i \>$,  with $i \in [1..n]$,  can be modeled with a MCS system $M=(C_1,\dots,C_n)$, where, for $i\in [1..n]$, $C_i(L,\ kb_i,\ ground(St(\MP_i)))$, $L$ is
		the ASP logic and $kb_i$ is obtained by removing the peer identifier from $(D_i \cup \LP_i \cup \IC_i)$.
		One can show that the set of maximal weak models of $\PS$ correspond to the minimal equilibria
		of the MCS obtained by removing from $M$ the bridge rules of diagnosis of the form $(D_1,\emptyset)$, where $D_1$ is minimal.
		A P2P system cannot be modeled by an MCS (in its basic form) if it contains minimal mapping rules.
		\begin{example}
			Let's consider the P2P system presented in Example \ref{Ex3-maximal}. It can be modeled by a MCS
			$M$ having two ASP contexts, $C_1=(L, kb_1,br_1)$ and $C_2=(L,kb_2,br_2)$, where $L$ is the ASP logic,
			$kb_1=\{\bot \leftarrow p(X),\ p(Y),\ X\neq Y\}$, $br_1=\{1:p(a)\leftarrow 2:q(a), 1:p(b)\leftarrow 2:q(b)\}$,
			$kb_2=\{q(a),\ q(b)\}$ and $br_2=\emptyset$. Clearly, $M$ is inconsistent because it does not admit any	equilibrium. Indeed, the atoms $p(a)$ and $p(b)$ are derived in $C_1$, its integrity
			constraint is violated and it does not have	an acceptable state.
			$M$ admits two minimal diagnosis of the form $(D_1,\emptyset)$.
			The first one is $(\{1:p(a)\leftarrow 2:q(a)\},\emptyset)$. If we remove its bridge rule from $M$, we obtain an MCS having only one minimal equilibrium: $(\{p(b)\},\{q(a),q(b)\})$. It corresponds to the maximal weak model $M_3$.\\
			The second one is $(\{1:p(b)\leftarrow 2:q(b)\},\emptyset)$ and removing its bridge rule from $M$, we obtain an MCS having the only minimal equilibrium $(\{p(a)\},\{q(a),q(b)\})$. It corresponds to the maximal weak model $M_2$.	\hfill $\Box$
		\end{example}
	}
	As for additional element of discussion, our proposal falls within the area of P2P system, 
	in which a generic peer is a kind of dynamic context whose presence is not guaranteed in the system, that is a peer may enter and leave the system, arbitrarily. Therefore, the focus in the P2P context (and also in our paper) is not that of finding the 
	explanations of inconsistencies, but just to cope with them.
	Moreover, in our work a generic peer is given the possibility to decide
	how to interact with a neighbor peer: 
	the use of maximal mapping rules states that it is preferable to import as long as no local inconsistencies arise; whereas the use of 
	minimal mapping rules states that it is preferable not to import unless a local inconsistency exists.\\
	This specific notion has not a counterpart in any of the above works in the field of MCS.

		\chr{
				In \cite{DBLP:journals/kais/BikakisAH11}  a fully distributed approach for reasoning in Ambient Intelligent Environments, based on the multi context system paradigm  has been proposed. 
			The paper refers to the propositional case and inconsistencies are managed 
			by prioritizing mapping rules that cause inconsistency, and specifically
			the decision which mapping rule to ignore is based, for every context, on the imposed strict total order of all contexts.
			Specifically, the user is forced to establish, at design time, the preference ordering on all the contexts and, as a consequence, 
			this allow to obtain a unique solution in polynomial time. 	
			As for a comparison,
			our approach models autonomous logic-based entities (peers) that interchange pieces of information
			using mapping rules.
			The essential feature of a P2P system is that each peer may leave and join the system arbitrarily. Due to this specific dynamic nature, our proposal avoids forcing any a priori preference ordering and as a consequence may admit many preferred weak models, whose computational complexity is in the second level of the polynomial hierarchy.
			In addition the work in \cite{DBLP:journals/kais/BikakisAH11} does not deal with the case in which the peer is locally inconsistent, whereas we can cope with this issue. 
			\chrr{	More generally, our proposal supports information flow between different agents through mapping rules, enables reasoning with inconsistent local information (minimal model semantics) and handles agents that provide mutually inconsistent information.  }
			On the other hand, it assumes that all peers share a common alphabet of constants and does not include any notion of privacy.
			In addition, the present proposal in its basic framework does not include any notion of preference between
			peers, which could be used to resolve potential conflicts caused by mutually inconsistent
			information sources and does not provide an algorithm for distributed computation.
			These last two features have been already investigated in other works of the same authors and have been briefly discussed in 
			Section \ref{Discussion}. 
			More specifically,  different extensions including preference criteria and aggregate functions have been proposed in \cite{ISMIS2008,IDEAS2011,FOIKS2012} and 
			a distributed computation  assigning semantics to a P2P system in polynomial time  has been presented in  \cite{adbis2017,iiwas2017}. \\
			Context theories can also be modeled as
			theories of defeasible logic \cite{DBLP:books/daglib/0013609,DBLP:books/daglib/0001812,Nute94},   mappings  as \emph{defeasible rules} and   
			a preference ordering on the system contexts is used to solve conflicts. 	
}

\section{Concluding Remarks and Directions for Further Research}\label{Concl}

In this paper we have proposed three different semantics 
for P2P deductive databases. 

In the \emph{Max Weak model Semantics} a peer imports maximal sets of atoms  from its neighborhood  to enrich its knowledge
while maintaining inconsistency anomalies.

In  the \emph{Min Weak model semantics} the  P2P system can be locally inconsistent, and the information provided by the neighbors is  used in order to restore
consistency, that is to only integrate with a missing portion of knowledge a correct but incomplete database.

In addition, the present paper unifies the previous two different
perspectives captured by the Maximal and Minimal Weak Model Semantics into  the 
\emph{Max-Min Weak Model Semantics}.
This declarative semantics, being more general, 
allows to characterize each peer in the neighborhood as a resource
used either to enrich (integrate) or to fix (repair) the knowledge,
so as to define a kind of \emph{integrate\-repair} strategy for
each peer in the P2P setting. 

\ch{
	The paper  also introduces
an alternative  characterization of the Max-Min Weak Model Semantics (resp. Max Weak Model Semantics and Min Weak Model Semantics)  by rewriting a P2P system into an equivalent prioritized logic program.
}
\chg{
Results on the complexity of answering queries are also presented. The paper,  by considering analogous results on stable model
semantics for prioritized logic programs, proves that for disjunction-free
($\vee-free$) prioritized programs
deciding whether an interpretation 
$M$ is a max-min weak model of $\PS$ is 
$co\NP$ complete; 
deciding whether
an atom is \emph{true} in some preferred model is
$\Sigma_2^p$-complete, whereas deciding whether an atom is
\emph{true} in every preferred model is $\Pi_2^p$-complete
\cite{SakIno00}. Moreover, the paper also provides results on the existence of a max-min weak model showing that the problem 
is in $\Sigma_2^p$.
}

\noindent 
Our work opens several avenues for future research. As a direction for further research, the work could be enriched by the introduction of preference criteria and explicit level of trusts so as to allow, in the presence of multiple
alternatives, the selection of data satisfying specific criteria and/or provided by the most reliable sources.

\nop{-------------------
In addition the proposed semantics follows the classical strategy of enriching as much as possible the knowledge of an information source, and preserves consistency by  minimally importing facts into a fixed set of relations. New and more sophisticated form of ...
----------}



\end{document}